\documentclass[aps,twocolumn,superscriptaddress,groupedaddress,floatfix]{revtex4}
\usepackage{amsmath,amsfonts}
\usepackage{graphicx}
\usepackage[pdftex,usenames,dvipsnames]{color}
\usepackage{comment}
\def\opone{\leavevmode\hbox{\small1\kern-3.8pt\normalsize1}}
\usepackage{subfigure}

\begin{document}

\title{Optical coherence and energy-level properties of a Tm$^{3+}$-doped LiNbO$_{3}$ waveguide at sub-Kelvin temperatures}

\author{Neil Sinclair}
\email[Corresponding author email: ]{neils@seas.harvard.edu}
\affiliation{Institute for Quantum Science and Technology, and Department of Physics \& Astronomy, University of Calgary, Calgary, Alberta T2N 1N4, Canada}
\affiliation{John A. Paulson School of Engineering and Applied Sciences, Harvard University, Cambridge, Massachusetts 02138, USA}
\affiliation{Division of Physics, Mathematics and Astronomy, and Alliance for Quantum Technologies (AQT), California Institute of Technology, Pasadena, California 91125, USA}

\author{Daniel Oblak}
\affiliation{Institute for Quantum Science and Technology, and Department of Physics \& Astronomy, University of Calgary, Calgary, Alberta T2N 1N4, Canada}

\author{Erhan Saglamyurek}
\affiliation{Institute for Quantum Science and Technology, and Department of Physics \& Astronomy, University of Calgary, Calgary, Alberta T2N 1N4, Canada}
\affiliation{Department of Physics, University of Alberta, Edmonton, Alberta, T6G 2E1, Canada}

\author{Rufus L. Cone}
\affiliation{Department of Physics, Montana State University, Bozeman, Montana 59717, USA}

\author{Charles W. Thiel}
\affiliation{Department of Physics, Montana State University, Bozeman, Montana 59717, USA}

\author{Wolfgang Tittel}
\affiliation{Institute for Quantum Science and Technology, and Department of Physics \& Astronomy, University of Calgary, Calgary, Alberta T2N 1N4, Canada}
\affiliation{QuTech and Kavli Institute of Nanoscience, Delft University of Technology, 2600 GA Delft, The Netherlands}

\begin{abstract}
We characterize the optical coherence and energy-level properties of the 795 nm $^3$H$_6$ to $^3$H$_4$ transition of Tm$^{3+}$ in a Ti$^{4+}$:LiNbO$_{3}$ waveguide at temperatures as low as 0.65 K. 
Coherence properties are measured with varied temperature, magnetic field, optical excitation power and wavelength, and measurement time-scale. 
We also investigate nuclear spin-induced hyperfine structure and population dynamics with varying magnetic field and laser excitation power. 
Except for accountable differences due to difference Ti$^{4+}$ and Tm$^{3+}$-doping concentrations, we find that the properties of Tm$^{3+}$:Ti$^{4+}$:LiNbO$_{3}$ produced by indiffusion doping are consistent with those of a bulk-doped Tm$^{3+}$:LiNbO$_{3}$ crystal measured under similar conditions.
Our results, which complement previous work in a narrower parameter space, support using rare-earth-ions for integrated optical and quantum signal processing.
\end{abstract}

\maketitle

\section{Introduction}
\label{sec:introduction}

Rare-earth-ion-doped crystals (REICs) cooled to cryogenic temperatures offer several exceptional properties for optical and radio-frequency processing tasks that range from amplifier development to quantum signal manipulation \cite{liu2006spectroscopic, macfarlane1987coherent}. 
These properties include desirable energy-level structures and population dynamics, long-lived optical and spin coherence, as well as their ability to be integrated on a chip \cite{thiel2011applications,tittel2010photon}. 
Of the many crystals that can host REIs, LiNbO$_{3}$ is attractive due to its transparency at optical wavelengths, high second-order nonlinearity, and ability to be modified to allow optical guiding, among other properties \cite{weis1985lithium}.
These attributes underpin the use of LiNbO$_{3}$ for several optical applications, including modulators that are used by the telecommunication industry \cite{wooten2000review}.

An early study of an ensemble of Tm$^{3+}$ ions in a Ti-indiffused LiNbO$_{3}$ waveguide (Tm$^{3+}$:Ti$^{4+}$:LiNbO$_{3}$) at 3 K demonstrated properties that were inferior to those measured in a Tm$^{3+}$-doped LiNbO$_{3}$ bulk crystal under similar conditions \cite{sinclair2010spectroscopic}. 
This difference was attributed to the perturbation of Tm$^{3+}$ ions by Ti$^{4+}$, which is used to raise the index of refraction of LiNbO$_{3}$ for waveguiding. 
However, a study of Tm$^{3+}$:Ti$^{4+}$:LiNbO$_{3}$ at 0.8 K revealed a tenfold improvement of properties compared to those measured at 3 K, matching those of a Tm$^{3+}$-doped bulk LiNbO$_{3}$ crystal under similar conditions \cite{sinclair2017properties}. 
Here we complement this study by measuring detailed coherence properties, population dynamics and sub-level structure of the $^3$H$_6$ to $^3$H$_4$ transition of a Tm$^{3+}$:Ti$^{4+}$:LiNbO$_{3}$ waveguide at temperatures as low as 0.65 K.

More precisely, using two- and three-pulse photon echo techniques, we measure coherence properties with varied laser excitation power and wavelength, temperature, magnetic field, and measurement time-scale \cite{liu2006spectroscopic, moerner1988persistent,macfarlane1987coherent}. 
We then characterize excited-level population dynamics and lifetimes using spectral hole burning. 
We also quantify nuclear spin-induced hyperfine energy level structure under the application of a magnetic field and with varying laser wavelength.
The hyperfine structure is probed further by coherent excitation and emission of light. 
Finally, we quantify the dependence of spectral hole widths and lifetimes on laser excitation power, both of which impact the suitability of Tm$^{3+}$:Ti$^{4+}$:LiNbO$_{3}$ for the aforementioned applications. 

Overall, we find equivalent properties to those measured using a 0.1\%-doped Tm$^{3+}$:LiNbO$_{3}$ bulk crystal at temperatures of less than 1 K except for accountable differences due to different Ti$^{4+}$- and Tm$^{3+}$-doping concentrations. 
Our results are relevant for integrated quantum and classical light processing, and clarify how properties of REIs are affected by crystal modification and measurement conditions in the technologically-significant LiNbO$_{3}$ material system.

\section{Experimental methods}
\label{sec:experimentalmethods}

Measurements are carried out using a 15.7 mm-long Tm$^{3+}$:Ti$^{4+}$:LiNbO$_{3}$ waveguide. 
It is created by raising the index of refraction of a $\sim$4 $\mu$m-wide strip by thermally indiffusing Ti into a 0.9 mm-thick, up to 0.7\% Tm$^{3+}$-indiffused, lithium niobate crystal wafer.
More details regarding the composition and fabrication of the waveguide can be found in Ref. \cite{sinclair2017properties}.
The crystal is mounted inside an adiabatic demagnetization refrigerator on an oxidized Cu stage and held down by a Macor lid that is attached to the Cu stage by spring-tensioned Ti screws. 
The Cu stage is mounted on an Au-coated Cu plate that is attached to a magnetically-shielded GGG paramagnetic salt pill that generates temperatures as low as 0.65 K by adiabatic demagnetization. 
The temperature is measured on the Macor lid using RuOx thermoelectric sensors.
Light is directed into, and out of, the waveguide by butt-coupling single-mode fiber at around 800 nm wavelength to the waveguide end facets. 
Each fiber is mounted on a three-axis nanopositioning stage to optimize the efficiency of the butt-coupling. 
Transmission through the entire cryogenic setup is 10-20\% due to imperfect overlap between the spatial modes of the fiber and the waveguide, reflections from uncoated surfaces, and imperfect fiber splices.
Magnetic fields of up to 20 kG are applied parallel to the c-axis of the crystal using a superconducting solenoid.
The magnetic field strength is determined using a Hall sensor mounted directly above the Macor lid.

We employ a continuous-wave external-cavity diode laser with an output power of up to 50 mW, an estimated linewidth of a few hundred kHz over millisecond timescales, and producing linearly-polarized light oriented approximately normal to the c-axis of the crystal (sigma polarization). 
The laser operating wavelength is tuned between 791 and 798 nm (vac.) by a diffraction grating that forms part of the laser cavity and monitored using a HeNe-referenced wavemeter featuring an accuracy better than 1 GHz. 
A 400 MHz acousto-optic modulator is used to produce pulses as short as 50 ns for photon echo measurements, or as long as 100 ms for hole burning and spectral tailoring of the absorption profile of the transition. 
Spectral features are probed by varying the laser detuning through serrodyne modulation using a standard 20 GHz-bandwidth LiNbO$_3$ phase modulator for frequency sweeps of $>20$ MHz.
A field-effect-transistor wired in parallel with the laser diode is used for frequency sweeps of $<20$ MHz. 
The laser power is varied from $\sim$1-10 mW for coherent transient (e.g. photon echo) measurements. 
It is kept at $\leq$1 mW for creating spectral features in order to avoid power broadening. 

Optical transmission is detected using a 1 GHz (2 MHz) AC (DC)-coupled amplified diode for photon echo (spectral tailoring-based) measurements and digitized using a 3 GHz digital oscilloscope. 
In cases where it is undesired, persistent spectral hole burning \cite{moerner1988persistent,macfarlane1987coherent} is mitigated by continuously varying the laser frequency over several GHz in tens of seconds using the diffraction grating. 
Moreover, we record photon echoes of largest intensities to further avoid spectral hole burning during photon echo excitation measurements.
For all other measurements, we average over many repetitions of the same experiment to minimize errors due to laser power fluctuations, frequency jitter, and noise.

\section{Results}
\label{sec:results}

\subsection{Optical coherence properties}
\label{sec:opticalcoherenceproperties}

Frequency-selective modification of the lineshape of an inhomogeneously-broadened transition of an ensemble of REIs is necessary for many optical processing applications \cite{tittel2010photon, thiel2011applications}. 
The maximum spectral resolution of the modification is ultimately determined by the homogeneous linewidth $\Gamma_h$ of the transition, which is inversely-proportional to the coherence lifetime $T_2= 1/\pi \Gamma_h$ \cite{mims1968phase}. 
Accordingly, decoherence can be interpreted as a broadening of the homogeneous linewidth \cite{macfarlane1987coherent}. 
This broadening is due to a time-varying detuning of the transition frequency of each REI due to dynamic perturbations caused by the lattice or neighboring ions \cite{liu2006spectroscopic}. 
These perturbations depend on the REIC and the experimental conditions used in the measurement. 
Thus, to better understand the nature of coherence in Tm$^{3+}$:Ti$^{4+}$:LiNbO$_{3}$, we quantify $\Gamma_h$ as a function of laser excitation power and wavelength, temperature, magnetic field, and measurement time scale (i.e. we perform time-dependent spectral diffusion).

\subsubsection{Wavelength dependence}
\label{sec:laserdetuningdependence}

\begin{figure}[ht!]
\begin{center}
\includegraphics[width=\columnwidth]{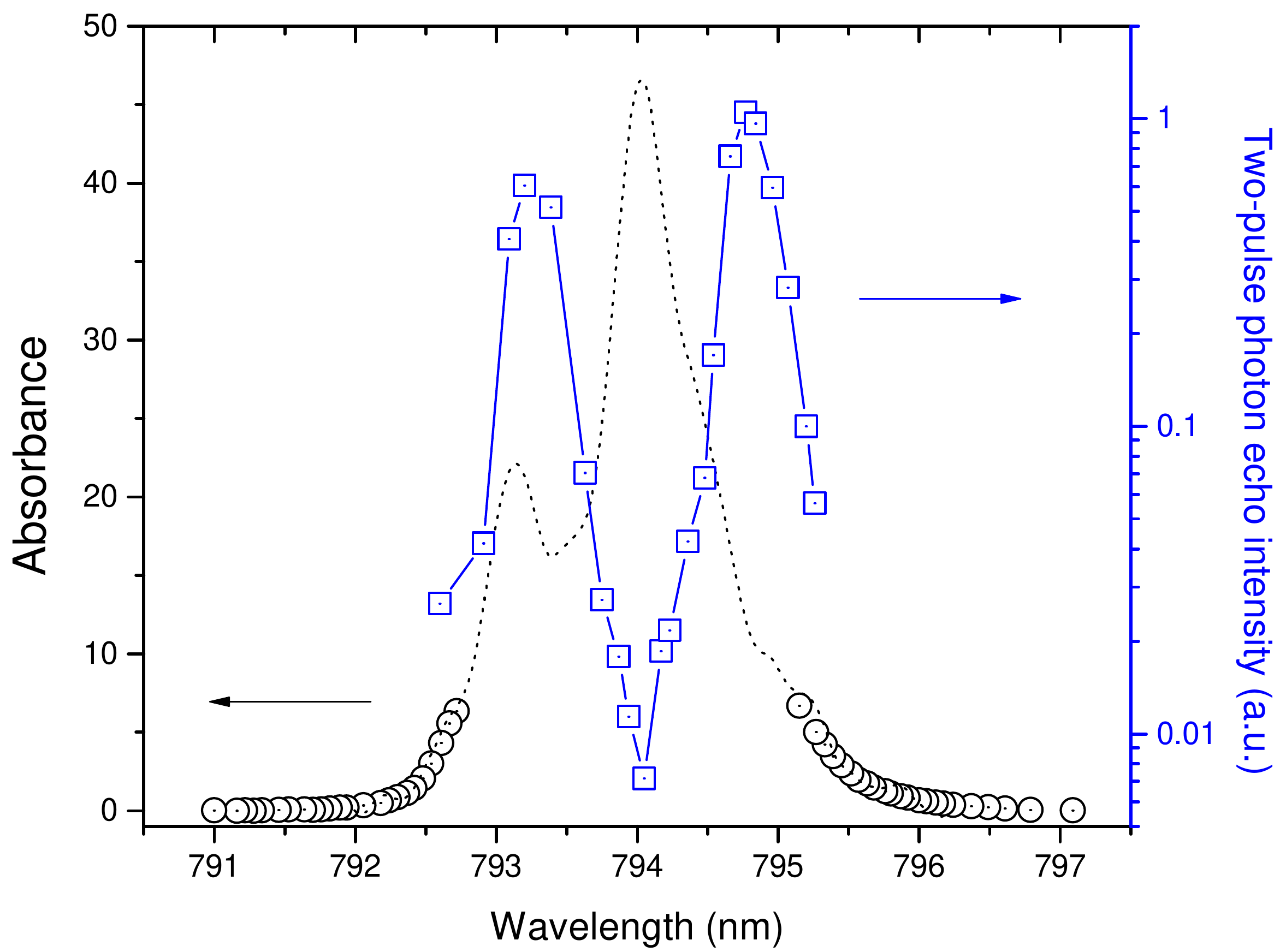}
\caption{Measured optical absorbance spectra from the  Tm$^{3+}$:Ti$^{4+}$:LiNbO$_{3}$ waveguide is shown using circles.
The dotted line indicates the absorption lineshape of a 0.1\%-doped Tm$^{3+}$:LiNbO$_{3}$ bulk crystal that has been vertically scaled by a factor of two.
Two-pulse photon echo intensity (log scale) measured from the waveguide is shown using squares with lines added to guide the eye.}
 \label{fig:inh_echo}
\end{center}
\end{figure}

The large inhomogeneous linewidth $\Gamma_{inh}$ of REIs in crystals is one of their unique features, which allows broadband or spectrally-multiplexed light-matter interactions \cite{macfarlane1987coherent, tittel2010photon}. 
Therefore, we quantify the inhomogeneous lineshape and probe optical coherence at various wavelengths to determine the bandwidth for which coherence properties are invariant. 
This lineshape may be determined using the Beer-Lambert relation $I_{out}=I_{off} e^{-d}$ in units of optical absorbance $d$. 
At 0.9 K, we generate a weak laser pulse, direct it into the waveguide, and detect its intensity $I_{out}(\lambda)$ at the output as a function wavelength $\lambda$ of the laser light. 
The laser is then tuned to an operating wavelength of 800 nm, off-resonant from the Tm$^{3+}$ transition, and the intensity $I_{off}$ of the weak pulse is recorded to yield the absorption lineshape (Fig. \ref{fig:inh_echo}, circles).
Note that the absorbance between 793 and 795 nm is too high to resolve the profile. 
This is expected due to the large oscillator strength of Tm$^{3+}$:Ti$^{4+}$:LiNbO$_{3}$, waveguide length, and doping concentration \cite{sinclair2010spectroscopic,thiel2010TmLN, sun2012TmLN, thiel2012waveguides}.
To compliment the measured data, Fig. \ref{fig:inh_echo} also shows a absorbance profile (dotted line) measured using a 0.1\%-doped Tm$^{3+}$:LiNbO$_{3}$ bulk crystal under similar conditions \cite{sun2012TmLN} that has been vertically scaled by a factor of two. 
This absorbance profile predicts a high $d>10$ between 793 and 795 nm for the waveguide while the scaling factor suggests an effective doping concentration of 0.2\%, which is consistent with the results of Ref. \cite{sinclair2010spectroscopic}.
Future measurements using a shorter sample or a lower Tm$^{3+}$ concentration will allow better characterization of the optical lineshape of Tm$^{3+}$:Ti$^{4+}$:LiNbO$_{3}$.
Nonetheless, our measurements unambiguously reveal $d<5$ for wavelengths between 795 and 797 nm, which is within the zero-phonon line \cite{macfarlane1987coherent,liu2006spectroscopic} and hence suitable for efficient and broadband signal processing. 

We now characterize wavelength-dependent decoherence at 0.9 K using two-pulse photon echos. 
Specifically, two pulses, each separated by a time duration of $t_{12}$, are directed into the waveguide and, due to the coherent response of the ions, a photon echo is produced a time $t_{12}$ later \cite{mims1968phase}. 
The intensity of the echo $I_{e}$ is described by \cite{sun2012TmLN, thiel2012echoOD} 
\begin{equation}
I_{e} \approx [e^{-d} \mbox{sinh}(d/2)]^2 e^{-4 t_{12}/ T_2}.
\label{eq:odecho}
\end{equation}
Re-absorption inhibits observation of an echo in the presence of high optical absorbance ($d>>1$). 
Therefore, using a 20 kG magnetic field, we generate a $\sim$300 MHz-wide spectral pit of reduced optical absorbance by optically pumping Tm$^{3+}$ ions to long-lived nuclear-hyperfine levels (see Sec. \ref{sec:nuclearzeeman}) \cite{macfarlane1987coherent,sinclair2016crossphase}. 
To avoid stimulated decay of population in the $^3$H$_4$ excited level, we wait 5 ms, which is an order of magnitude longer than the $\sim$ 100 $\mu$s lifetime of this level (see Sec. \ref{sec:lifetimeexcited}), and perform two pulse photon echo excitation for a fixed $t_{12}$ and varied wavelength to determine $I_{e}(\lambda)$ (Fig. \ref{fig:inh_echo}, squares). 

Our optical pumping sequence is performed with maximum laser power for all measurements, resulting in a varying residual absorbance as a function of wavelength at the bottom of the spectral pit.
Yet this absorbance is still high ($d>>1$) for wavelengths around 794 nm where we expect maximum absorbance.
Accordingly, $I_{e}$ is reduced for wavelengths around 794 nm likely due to re-absorption rather than reduced $T_2$ \cite{sun2012TmLN}, an effect that is not taken into account in Eq. \ref{eq:odecho}.
Consequently, we are unable to accurately determine $T_2(\lambda)$ using Eq. \ref{eq:odecho}.
Note that $T_2$ was found to be invariant over $\sim$ 300 GHz of bandwidth around 794.5 nm using bulk Tm$^{3+}$:LiNbO$_{3}$ at 1.6 K \cite{sun2012TmLN}. 
Nonetheless, we observe photon echoes over $>$ 100 GHz of bandwidth, suggesting coherence over this range.

As a last step, we vary the laser power and detuning to generate a spectral population grating \cite{macfarlane1987coherent, sinclair2017properties} instead of a spectral pit.
After a time delay of 5 ms, we generate a laser pulse that scatters from the grating to produce an echo 200 ns later. 
This protocol is similar to three-pulse photon echo excitation (see Sec. \ref{sec:spectraldiffusionwithfield}) \cite{macfarlane1987coherent}. 
We repeat this procedure at various wavelengths, finding that the intensity of the echo varies with wavelength similar to that of the two-pulse photon echos shown in Fig. \ref{fig:inh_echo}.
This suggests that the coherence properties along with the underlying level structure and dynamics are suitable for wide-band optical processing.

\subsubsection{Temperature dependence}
\label{sec:temperaturedependence}

\begin{figure}[ht!]
\begin{center}
\includegraphics[width=\columnwidth]{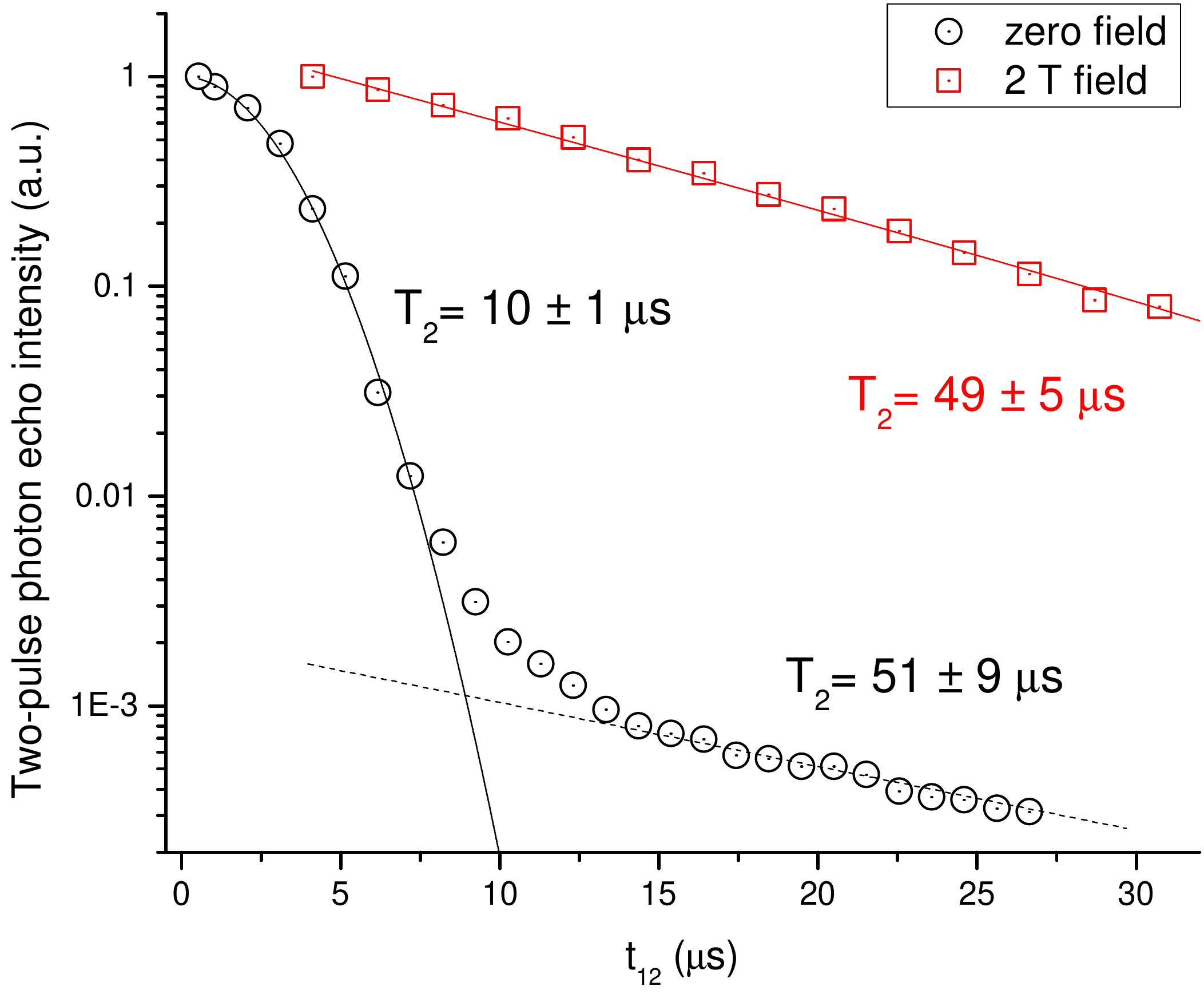}
\caption{Example two-pulse photon echo decays using zero (circles) and 20 kG (squares) magnetic field on a log scale.
All data is taken at 0.89 K with excitation pulse powers of $\sim$ 1 mW. 
The black solid line corresponds to the fit using Eq. \ref{eq:mims}, while the black dotted line indicates the motional-narrowing regime.
The red line is the fit of Eq. \ref{eq:mims} to the 20 kG data.} 
 \label{fig:t12decays}
\end{center}
\end{figure}

\begin{figure}[ht!]
\begin{center}
\includegraphics[width=\columnwidth]{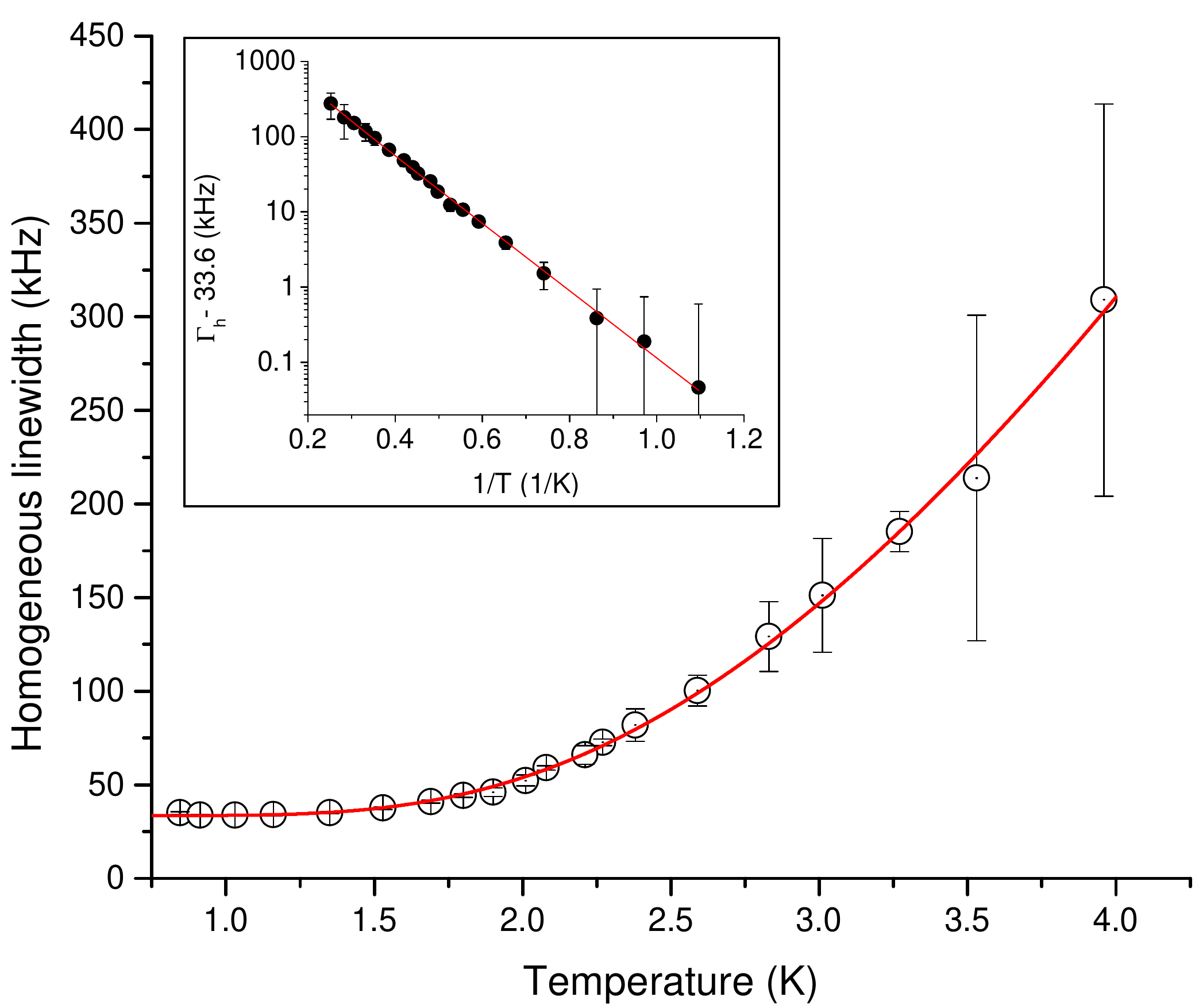}
\caption{Temperature dependence of homogeneous linewidth measured using two-pulse photon echoes. 
A fit of the data using Eq. \ref{eq:direct} is shown by the red line.
Inset: data and fit shown in the main figure (log scale) against $1/T$ to highlight the exponential dependence of $\Gamma_h-\Gamma_0$.}
 \label{fig:gh_temp_B0}
 \end{center}
\end{figure}

Phonon interactions are a fundamental cause of decoherence of REIs \cite{liu2006spectroscopic}. 
Therefore, temperature-dependent measurements are critical for determining the limitations of coherence properties of REICs. 
Here we tune the laser operating wavelength to 795.0 nm and measure $\Gamma_h$ as a function of temperature by measuring the decay of two-pulse photon echo intensities as the time delay $t_{12}$ between the two excitation pulses is varied. 
This wavelength is used to ensure strong light-matter interaction without the reduction of signal intensity due to reabsorption.
This and similar wavelengths are also used for several other measurements in this work (e.g. photon echo excitation and hole burning) for the same reasons.
The echo intensity $I$ is measured and fit using the Mims expression
\begin{equation}
I(t_{12}) = I_0 e^{-2(\tfrac{2 t_{12}}{T_2} )^x},
\label{eq:mims}
\end{equation}
where $I_0$ is the echo intensity at $t_{12}=0$, and $x$ is an empirical parameter determined by spectral diffusion (also discussed in Secs. \ref{sec:magneticfielddependence} and \ref{sec:spectraldiffusionwithfield}) \cite{mims1968phase}. 
At 4 K the decay is exponential (i.e. $x=1$), indicating that it is dominated by phonon scattering \cite{liu2006spectroscopic,sun2012TmLN}, with $\Gamma_h \approx 300$ kHz.
However, at lower temperatures the decay becomes non-exponential (Fig. \ref{fig:t12decays}, circles).
The initial part of the decay shows a time-dependent increase of decoherence due to spectral diffusion (indicated by a black line in Fig. \ref{fig:t12decays}), while the later part of the decay becomes again exponential -- a characteristic of motional narrowing (dotted line in Fig. \ref{fig:t12decays}) \cite{liu2006spectroscopic, thiel2010TmLN, sun2012TmLN}. 
The motional narrowing regime, in which the accumulated dephasing is reduced due to spins flipping back to their original state, is distinguished by the lack of time-dependent spectral diffusion.
The time at which motional narrowing appears suggests a spectral diffusion rate of $\sim$ 100 kHz, which is similar to that measured using bulk Tm$^{3+}$:LiNbO$_{3}$ at 1.7 K and a wavelength of 794.26 nm \cite{thiel2010TmLN, sun2012TmLN}. 
The decay for $t_{12}<7$ $\mu$s is used for the temperature-dependent characterization and, from fits to Eq. \ref{eq:mims}, the temperature-dependent homogeneous linewidth is determined (Fig. \ref{fig:gh_temp_B0}).

We expect $\Gamma_h$ to be limited by direct phonon excitation to a higher-lying crystal field level in the ground state manifold \cite{liu2006spectroscopic, sun2012TmLN}:
\begin{equation}
\Gamma_h = \Gamma_{0} + \frac{\Gamma_{ph}}{e^{\Delta E /k_{B}T} -1},
\label{eq:direct}
\end{equation}
where $\Gamma_{0}$ is the homogeneous linewidth at zero temperature, $\Gamma_{ph}$ is the phonon coupling coefficient, $\Delta E$ is the phonon energy, $T$ is temperature, and $k_{B}$ is the Boltzmann constant. 
A fit of Eq. \ref{eq:direct} to the data in Fig. \ref{fig:gh_temp_B0}, yields an intrinsic linewidth of $33.5\pm 1.5$ kHz, agreeing with the 30 kHz predicted from measurements of a bulk Tm$^{3+}$:LiNbO$_{3}$ crystal at a wavelength of 794.26 nm \cite{sun2012TmLN}. 
We also find $\Delta E = 7.0 \pm 0.2$ cm$^{-1}$ and $\Gamma_{ph} = 3.2 \pm 0.3$ MHz. 
The former agrees well with the ground-state crystal field splitting of 7.2 cm$^{-1}$ of bulk Tm$^{3+}$:LiNbO$_{3}$ while the latter is three times higher than the 1.1 MHz observed using the bulk crystal at 795.01 nm \cite{sun2012TmLN}.
This supports our previous observation of non-ideal coherence properties of Tm$^{3+}$:Ti$^{4+}$:LiNbO$_{3}$ compared to bulk Tm$^{3+}$:LiNbO$_{3}$ at 3 K \cite{sinclair2010spectroscopic}.
The large phonon coupling may be due to the higher, up to 0.7\%, doping concentration of Tm$^{3+}$:Ti$^{4+}$:LiNbO$_{3}$ compared to 0.1\%-doped bulk Tm$^{3+}$:LiNbO$_{3}$ \cite{thiel2019private}.
Since $\Delta E \gg kT$ for all of our measurements, $\Gamma_h -\Gamma_0$ exhibits an exponential dependence with respect to $1/T$ (Fig. \ref{fig:gh_temp_B0} inset, the fit using Eq. \ref{eq:direct} is also shown). 

\subsubsection{Magnetic field dependence}
\label{sec:magneticfielddependence}

Decoherence beyond that induced by direct phonon-ion interactions is due to spectral diffusion. 
This may be caused by fluctuating fields within the host crystal that are generated by dynamic interactions between host spins or impurities \cite{liu2006spectroscopic}. 
Spectral diffusion results in a broadening of the measured homogeneous linewidth because each ion experiences a slightly-different dynamic environment.
A magnetic field is expected to reduce the impact of spectral diffusion by increasing the energy-splitting between magnetic levels beyond the phonon energy, inducing spin polarization and reducing the number of spin flips \cite{liu2006spectroscopic}. 
Furthermore, the applied field reduces decoherence through the 'frozen core' effect in which the large magnetic moment of REIs create a localized magnetic field gradient that inhibits spin flips of nearby nuclear spins, creating a spin diffusion barrier \cite{shelby1978optically}. 
Spectral diffusion is expected due to coupling between Tm$^{3+}$ and $^{93}$Nb, $^7$Li, other Tm$^{3+}$, and possibly Ti$^{4+}$ nuclear spins in Tm$^{3+}$:Ti$^{4+}$:LiNbO$_3$ \cite{thiel2010TmLN, sun2012TmLN,thiel2012waveguides}.

We apply a magnetic field of 20 kG and measure a two-pulse photon echo decay at a temperature of 0.89 K and wavelength of 795.0 nm. 
We find an exponential decay that is free of time-dependent spectral diffusion (Fig. \ref{fig:t12decays}, squares), consistent with Tm$^{3+}$:LiNbO$_{3}$ bulk crystal measurements at a wavelength of 794.26 nm \cite{thiel2010TmLN, sun2012TmLN}.
Moreover, fitting this decay using Eq. \ref{eq:mims} reveals a coherence lifetime of $49 \pm 5$ $\mu$s, which is consistent with the $51 \pm 9$ $\mu$s observed in the motional narrowing regime of the zero field data.
We attribute the reduction of coherence lifetimes compared to the 117 $\mu$s reported in Ref. \cite{sinclair2017properties} to non-zero excitation-induced decoherence during our measurement (see Sec. \ref{sec:excitationinducedspectraldiffusion} for an analysis of this effect).

\begin{figure}[ht!]
\begin{center}
\includegraphics[width=\columnwidth]{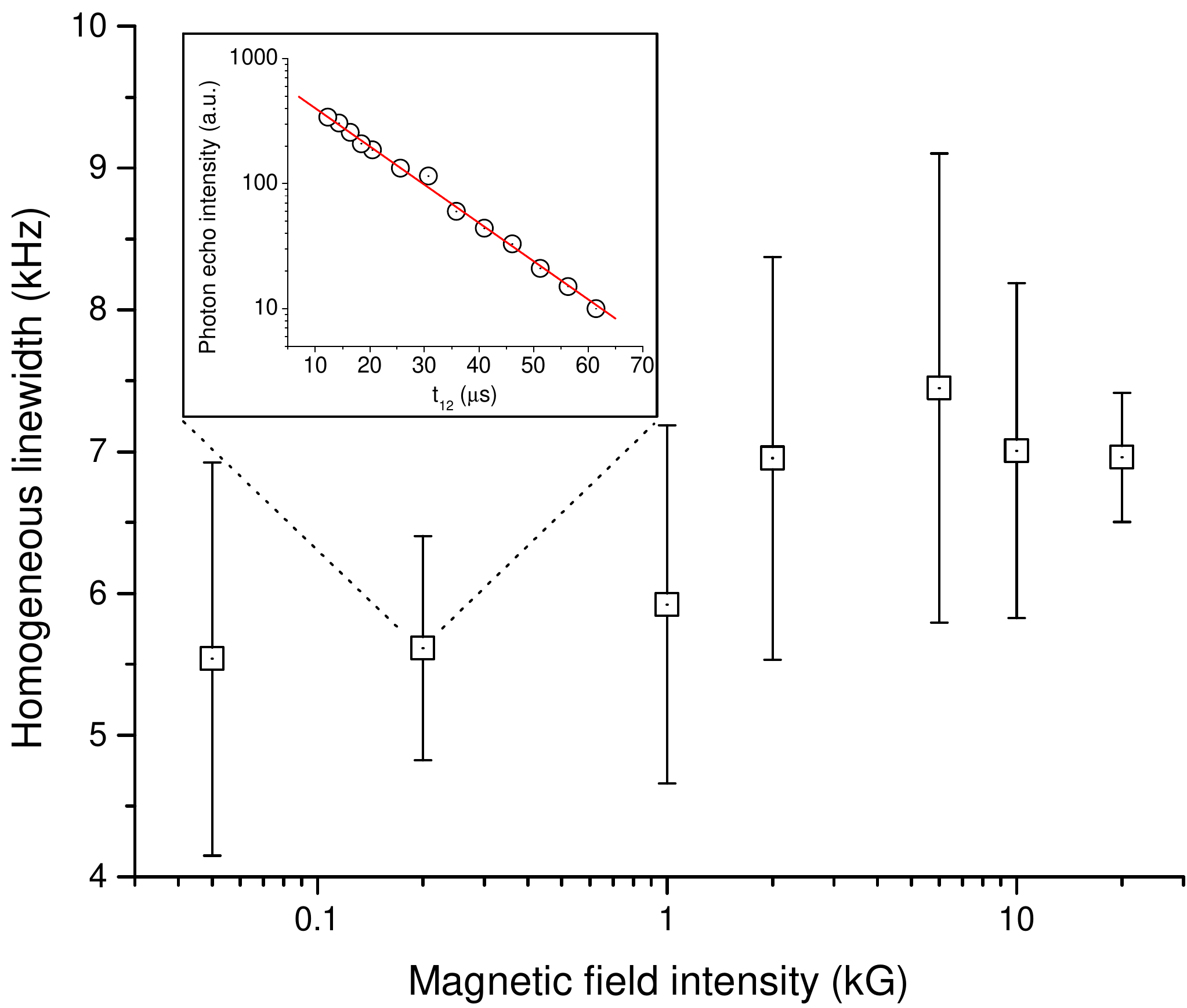}
\caption{Magnetic field dependence of $\Gamma_h$ (linear-log scale). 
Inset: two-pulse photon echo decay and fit (log scale) using a 200 G magnetic field.} 
 \label{fig:gh_Bfield}
\end{center}
\end{figure}

To further investigate magnetic field-dependent decoherence, we measure two-pulse photon echo decays for varying magnetic fields at a temperature of 0.86 K and fit the decays using Eq. \ref{eq:mims} to determine $\Gamma_h$ (Fig. \ref{fig:gh_Bfield}).
For reliable fits, we perform echo decays for $t_{12}>5 \mu$s when echo modulation has diminished (see Sec. \ref{sec:superhyperfine}).
We find little field dependence and that 100 G of field is enough to achieve coherence properties comparable to that of the zero-field motional narrowing regime.
Note that we observe a similar field dependence when performing the measurement using three times less excitation power in order to avoid the impact of excitation-induced spectral diffusion. 

\subsubsection{Excitation-induced spectral diffusion}
\label{sec:excitationinducedspectraldiffusion}

\begin{figure}[ht!]
\centering
\includegraphics[width=\columnwidth]{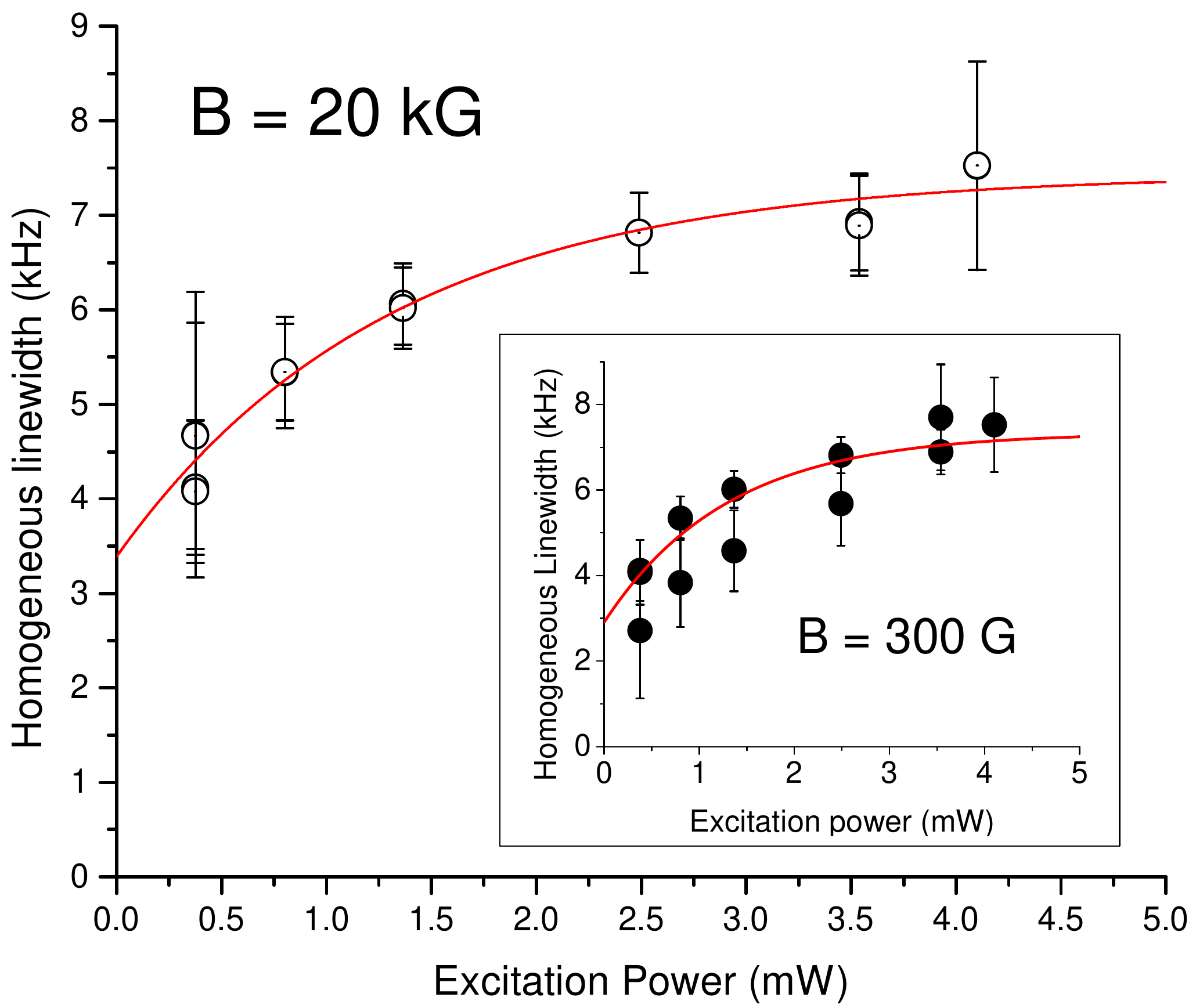}
\caption{Excitation power-dependence of the homogeneous linewidth. 
Measurements using a 20 kG (300 G) field is shown in the main figure (inset). 
Excitation power refers to the peak power of the second excitation pulse at the beginning of the waveguide, accounting for any insertion loss.
} 
\label{fig:ISD_bothfields}
\end{figure}

One source of spectral diffusion is due to optical excitation. 
Its effect on reducing coherence lifetimes is referred to as instantaneous spectral diffusion (ISD) \cite{liu1987photon,thiel2014ISD}. 
As REIs are excited, the change in their permanent electric dipole moment perturbs the resonance frequencies of neighboring ions via electric dipole interactions.
Consequently, the coherence lifetime that is inferred from a two-pulse photon echo decay may be underestimated if intense excitation pulses are employed. 
To determine the limitations of the coherence lifetime due to ISD, we set the temperature to be between 0.81 and 0.89 K and measure two-pulse photon echo decays at a wavelength of 795.6 nm with varying excitation powers and magnetic fields of 300 G and 20 kG. 
The echo decays are fit using Eq. \ref{eq:mims}, and we observe that the homogeneous linewidth rises with excitation power until saturation (Fig. \ref{fig:ISD_bothfields}). 
This behavior is supported by the following rate equation model for ISD:
\begin{equation}
\Gamma_{h}=\Gamma_{P=0}+\frac{1}{2}\Gamma_{ISD}(1-e^{-c_P P}),
\end{equation}
where $\Gamma_{P=0}$ is the homogeneous linewidth at zero excitation power, $\Gamma_{ISD}$ is the full-width at half maximum spectral broadening due to ISD and $c_P$ is a constant \cite{ thiel2014ISD}. 
A fit of our data using this model yields $\Gamma_{P=0}=2.9 \pm 1.1$ kHz ($3.4 \pm 0.3$ kHz) and $\Gamma_{ISD}=8.9 \pm 1.7$ kHz ($8.1 \pm 0.5$ kHz) for the measurements at 300 G (20 kG) field, consistent with values reported in Ref. \cite{ thiel2014ISD}. 
The constant $c_P$ is related to a parameter $\beta_{ISD}$ that only depends on the strength and the nature of the ion-ion interaction. 
It is intrinsic to the material. 
For our measurements, $\beta_{ISD}$ may be estimated using
\begin{equation}
\beta_{ISD}= \frac{3 \pi {\omega_0}^2 h c}{2 \lambda \alpha t_p} c_P \Gamma_{ISD},
\end{equation}
where $\omega_0$ is the radius of waveguide mode, $\lambda$ is the optical excitation wavelength, $\alpha$ is the absorption coefficient and $t_p$ is the duration of the excitation pulse \cite{thiel2014ISD}. 
The waveguide features $\omega_0=6.25$ $\mu$m and $\alpha=1.1$ cm$^{-1}$ at $\lambda=795.6$ nm, and we employ excitation ($\pi$-)pulses of duration $t_p=25.6$ ns.
This yields $\beta_{ISD}=5.1 \pm 3.2 \times 10^{-11}$ Hz$\cdot$cm$^{3}$ ($5.1 \pm 1.1 \times 10^{-11}$ Hz$\cdot$cm$^{3}$) for the 300 G (20 kG) data, which is consistent with the $6 \times 10^{-11}$ Hz$\cdot$cm$^{3}$ measured using bulk Tm$^{3+}$:LiNbO$_{3}$ at 1.7 K and a wavelength of 794.27 nm \cite{thiel2014ISD}. 
Note that the effects of ISD could not be quantified at zero field due to a weak photon echo intensity. 
Furthermore, varying amounts of ISD are measured at different wavelengths around 795.6 and 794.2 nm, consistent with observations using bulk Tm$^{3+}$:LiNbO$_{3}$ \cite{thiel2019private}.
This suggests more complicated ion-ion or spin-spin interactions that go beyond the model discussed here \cite{thiel2012waveguides,thiel2014ISD}, and should be addressed in future work.

\subsubsection{Time-dependent spectral diffusion with magnetic field}
\label{sec:spectraldiffusionwithfield}

In the presence of a magnetic field, spectral diffusion is known to occur over timescales longer than the maximum value of $t_{12}=$ 50 $\mu$s used for our two pulse photon echo excitation measurements \cite{liu2006spectroscopic,macfarlane1987coherent}. 
Since many optical signal processing applications using REIs rely on spectral features being created and probed over long timescales \cite{thiel2011applications, tittel2010photon,macfarlane1987coherent}, we study spectral diffusion on such timescales.
Towards this end, we use three-pulse photon echos at magnetic fields of 300 G and 20 kG. 
Specifically, we generate two pulses that are separated by a time duration of $t_{12}$ to create a population grating. 
Here, this grating is formed by optical pumping and trapping of population in the $^3$H$_4$, nuclear-hyperfine, or superhyperfine levels (see Secs. \ref{sec:lifetimeexcited}, \ref{sec:nuclearzeeman}, and \ref{sec:superhyperfine}). 
A third pulse, generated $t_{23}$ after the second pulse, is scattered from the grating to produce a photon echo of intensity
\begin{equation}
I(t_{23})=I_0 {I_{pop}}^2(t_{23}) e^{-4 t_{12} \Gamma_{h}(t_{23})},
\label{eq:3pe}
\end{equation}
where $\Gamma_{h}(t_{23})$ is the time-dependent homogeneous linewidth, $I_0$ is a normalization constant, and $I_{pop}(t_{23})$ represents the reduction in echo intensity due to population decay \cite{ bottger2006ErYSO}. 
For Tm$^{3+}$:Ti$^{4+}$:LiNbO$_{3}$ at low temperatures and under magnetic fields, $I_{pop}(t_{23}) \approx C_{1}e^{-t_{23}/T_1}+ C_{B}e^{-t_{23}/T_B}+C_{H}e^{-t_{23}/T_H}$,
where each $C_i$ are constants and $T_1=109$ $\mu$s is the population lifetime of the $^3$H$_4$ level, $T_B=2.9$ ms is the bottleneck level population lifetime (see Sec. \ref{sec:lifetimeexcited}), and $T_H>1$ s is the population lifetime of the hyperfine levels (see Secs. \ref{sec:nuclearzeeman} and \ref{sec:superhyperfine}). 
All lifetimes are much longer than $t_{12}$.

\begin{figure}[ht!]
\begin{center}
\includegraphics[width=\columnwidth]{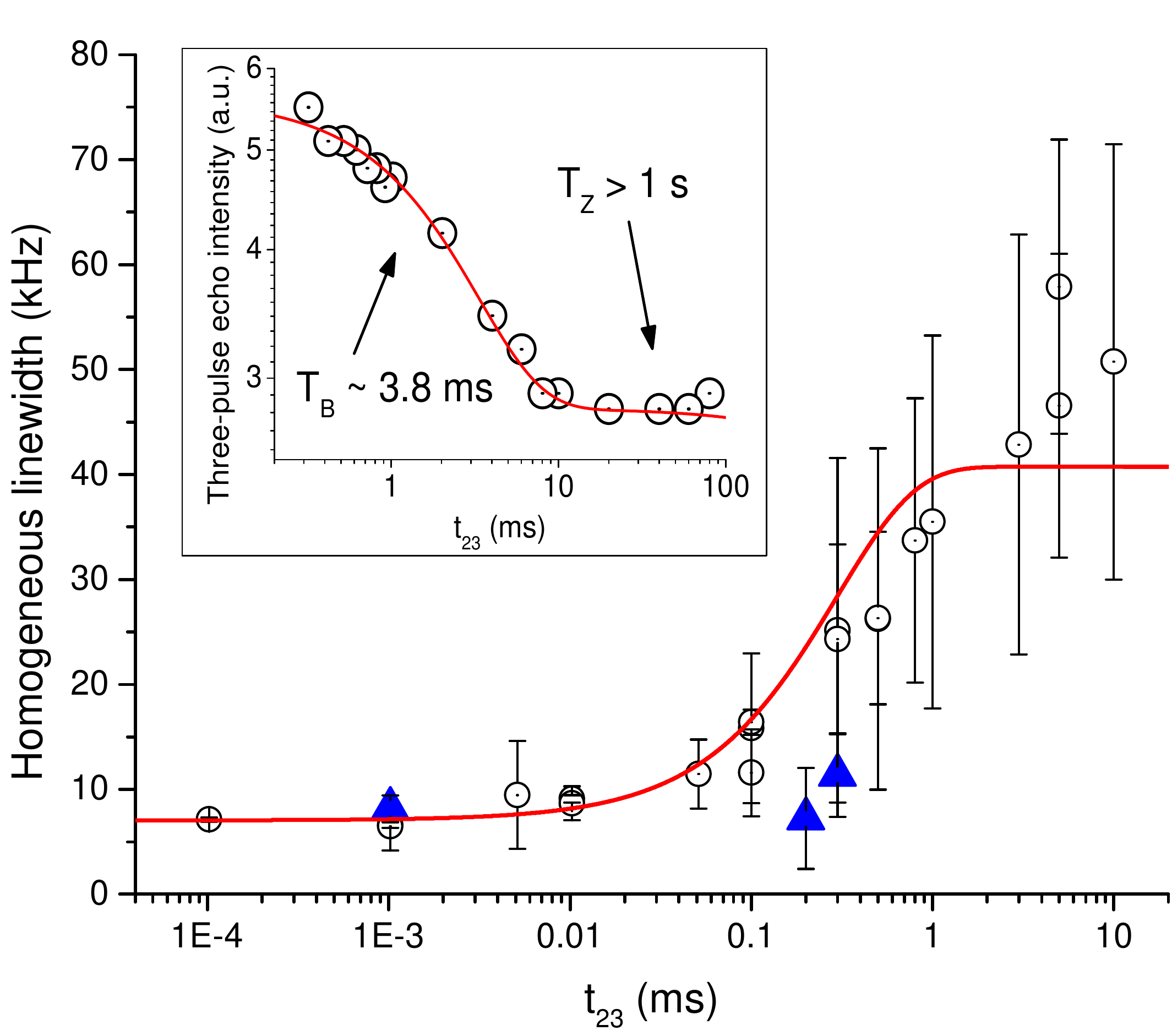}
\caption{Time-dependent decoherence revealed by three-pulse photon echoes. 
Variation of the homogeneous linewidth with $t_{23}$ (linear-log scale) at 300 G (20 kG) magnetic field is shown using triangles (circles). 
Note that the increasing uncertainty of $\Gamma_h$ with $t_{23}$ is due to decreasing echo intensities as a result of the decay of the $^3$H$_4$ level (see Sec. \ref{sec:lifetimeexcited}) and the restriction of echo decays to timescales beyond which echo modulation has subsided ($t_{12}>500$ ns, see Sec. \ref{sec:superhyperfine}).
Inset: variation of the three pulse photon echo intensity with $t_{23}$ for $t_{23}> T_1$ (double-log scale).} 
 \label{fig:gh_t23}
\end{center}
\end{figure}

Specifically, at temperature of 0.80 K, a wavelength of 795.3 nm, and using a magnetic field of 300 G, we vary $t_{12}$, and fit the echo decay using Eq. \ref{eq:mims} ($x=1$) to determine $\Gamma_h (t_{23})$ up to $t_{23}= 300$ $\mu$s (Fig. \ref{fig:gh_t23}, triangles).
We find that spectral diffusion is negligible over the entire measurement timescale. 
An $\sim$8 kHz homogeneous linewidth is observed, which is consistent with the magnetic field-dependent results presented in Fig. \ref{fig:gh_Bfield} given the strong ISD present during the measurement, and represents a slight improvement over the $\sim$10 kHz measured using bulk Tm$^{3+}$:LiNbO$_{3}$ at 1.7 K and 794.26 nm wavelength, perhaps due to lowered temperature \cite{sun2012TmLN, thiel2010TmLN, thiel2012waveguides}.

Next, we increase the field to 20 kG and repeat the measurement at 0.95 K (Fig. \ref{fig:gh_t23}, circles).
We find a linewidth that grows to $\sim$50 kHz after 10 ms, indicating a small rate of nuclear spin flips in the host, which may be due to the frozen core effect.

To quantify the processes driving spectral diffusion, we fit our data using a spin fluctuation model
\begin{equation}
\Gamma_{h} (t_{23})= \Gamma_0 +\frac{1}{2}\Gamma_{SD} (1-e^{-R_{SD} t_{23}}),
\label{eq:sd}
\end{equation}
where $\Gamma_0$ is the homogeneous linewidth at $t_{23}=0$, and $\Gamma_{SD}$ is the maximum linewidth broadening due to spin fluctuations at a rate $R_{SD}$ \cite{mims1968phase,  bottger2006ErYSO,klauder1962spectral,mims1972spect}. 
We find $\Gamma_0=7.0\pm0.2$ kHz, which is consistent with the magnetic field- and excitation power-dependence of $\Gamma_h$ shown in Figs. \ref{fig:gh_Bfield} and \ref{fig:ISD_bothfields}, respectively.
Our fit also yields $\Gamma_{SD}=67.4\pm12.9$ kHz, which is a factor of two larger than that measured using bulk Tm$^{3+}$:LiNbO$_3$ with zero magnetic field at 1.7 K and 794.26 nm wavelength \cite{sun2012TmLN}. 
Moreover, we also find $R_{SD}=3.4\pm0.9$ kHz, which is nearly 40 times smaller than that measured using the aforementioned bulk crystal and experimental conditions \cite{sun2012TmLN}. 
While the reduced spectral diffusion rate is probably due to differences in field and temperature compared to those used for the measurements of bulk Tm$^{3+}$:LiNbO$_{3}$, the increased maximum broadening may be due to additional (and not yet understood) dynamics under strong fields, at long timescales, and at this wavelength, which will be explored in future studies.

Note that despite the strong excitation powers used in this measurement, we do not observe the distinct increase (decrease) in decoherence over 1 $\mu$s $<t_{23}<$ 10 $\mu$s (at $t_{23} \sim$ 110 $\mu$s) that was observed using bulk Tm$^{3+}$:LiNbO$_3$ at a wavelength of 794.27 nm and zero magnetic field \cite{thiel2012waveguides, thiel2014ISD}.
This decoherence was ascribed to the magnetic character of the $^3$H$_4$ multiplet and population decay of these levels.
The strong magnetic field and low temperature used here likely inhibits these magnetic interactions by spin polarization.
%model for t12=0 suitable here because decays are exponential. 

To examine the effects of spectral diffusion beyond timescales of 10 ms, we perform three pulse photon echo excitation and measure the echo intensity with $t_{23}$ varied from 0.2 to 100 ms for fixed $t_{12}=500$ ns (Fig. \ref{fig:gh_t23} inset). 
A fit using Eq. \ref{eq:3pe} reveals that the echo intensity depends only on $I_{pop}$ with $C_1 \approx 0$ (since $t_{23}>T_1$), $T_B \approx 3.8$ ms and $T_H>1$ s. 
Our fit suggest that for $t_{23}>10$ ms, $\Gamma_h$ remains constant and any reduction in echo intensity is solely due to population decay from the hyperfine levels. 
This is consistent with the model of Eq. \ref{eq:sd} to describe the spectral diffusion dynamics in which the homogeneous linewidth saturates at long time delays.

\subsection{Population dynamics and energy level structure}
\label{sec:populationdynamics}

In addition to long coherence lifetimes of the optical transition, long-lived energy levels are required to realize optical technology with REIs \cite{liu2006spectroscopic, macfarlane1987coherent, thiel2011applications}.
For instance, these levels are used as a population reservoir for spectral tailoring of diffraction gratings or quantum memories.
Therefore, we characterize both the energy-level structure and population dynamics of the $^3$H$_6$ to $^3$H$_4$ transition of the Tm$^{3+}$:Ti$^{4+}$:LiNbO$_{3}$ waveguide under varying conditions.

\subsubsection{Lifetimes and dynamics of excited levels}
\label{sec:lifetimeexcited}

\begin{figure}[ht!]
\begin{center}
\includegraphics[width=\columnwidth]{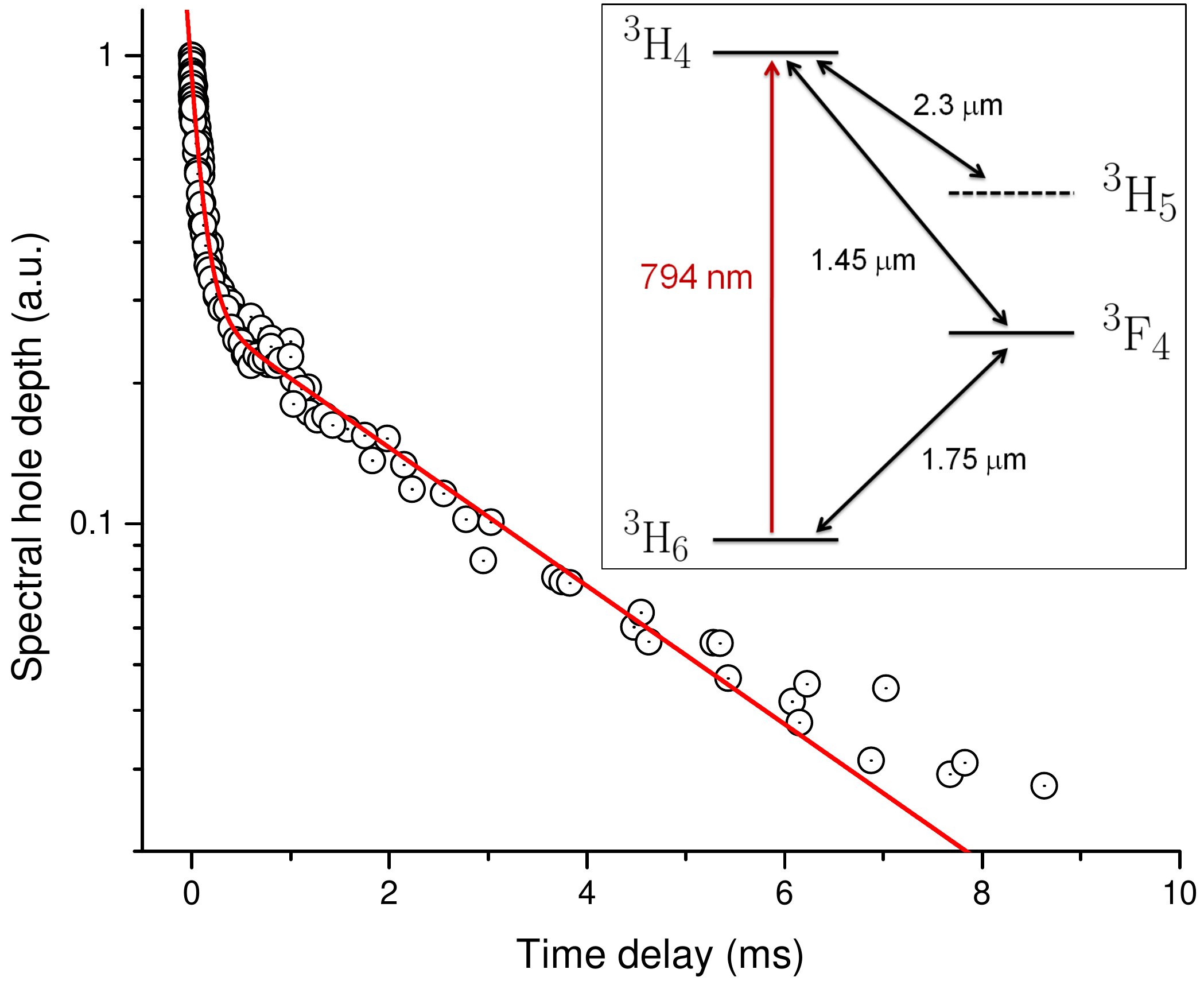}
\caption{Transient spectral hole decay and fit (log scale). 
Inset: Simplified electronic level structure and approximate transition wavelengths of Tm$^{3+}$. 
} 
 \label{fig:T1}
\end{center}
\end{figure}

We perform time-resolved spectral hole burning, in which interaction of a short laser pulse excites a subset of ions and, after a varying time delay $t_{d}$, the resultant increase in optical transmission is assessed by varying the laser frequency and measuring the depth of the spectral hole \cite{moerner1988persistent, macfarlane1987coherent}. 
We aim to determine the population lifetimes of the $^3$H$_4$ excited level ($T_1$) and the $^3$F$_4$ `bottleneck' level ($T_B$) as shown in the simplified energy level diagram of the inset of Fig. \ref{fig:T1} \cite{liu2006spectroscopic}.
The $^3$H$_5$ level has a lifetime much shorter than $T_B$ and cannot be observed in the hole decay.
Our measurements are performed at zero field, a temperature of 0.85 K, and at a wavelength of 795.50 nm.
Note that, in accordance with the results of Ref. \cite{thiel2012waveguides}, measurements at this wavelength probe ions that experience more local strain or those that occupy multiple sites.

The variation of spectral hole depth with time delay reveals the expected double-exponential (Fig. \ref{fig:T1}). 
Note that the hole depth is proportional to the number of excited ions because spectral diffusion has saturated after 10 $\mu$s. 
A fit of the decay using
\begin{equation}
e^{-t_{23}/T_1}+\frac{\beta}{2}\frac{T_B}{T_B-T_1}(e^{-t_{23}/T_B}-e^{-t_{23}/T_1}),
\end{equation}
where $\beta$ is the branching ratio to the bottleneck level \cite{sun2012TmLN}, revealing $T_1 = 109\pm 7$ $\mu$s, $T_B = 2.9\pm 0.4$ ms, and a branching ratio $\beta = 62\pm 3\%$. 

We find a difference in lifetimes and branching ratio to those measured previously at 795.52 nm and 3.5 K, where $T_1 = 82$ $\mu$s, $\beta = 44\%$, and $T_B = 2.4$ ms is reported \cite{sinclair2010spectroscopic}. 
We attribute this difference to the possibility of addressing different subsets of ions \cite{sun2012TmLN, thiel2012waveguides} or a very weak signal-to-noise ratio that distorted the previous results. 
Comparing our measured lifetimes with those obtained using bulk Tm$^{3+}$:LiNbO$_{3}$ \cite{thiel2010TmLN}, where $T_1 = 160$ $\mu$s, $\beta = 27\%$, and $T_B = 4.5$ ms was measured at a wavelength of 794.28 nm, we find shorter lifetimes and a larger branching ratio. 
This could again be due to the difference in wavelength or, more likely, to a larger doping concentration of Tm$^{3+}$:Ti$^{4+}$:LiNbO$_{3}$.
Increased doping enhances the Tm$^{3+}$-Tm$^{3+}$ cross-relaxation rate via decay through the $^3$H$_5$ and $^3$F$_4$ level \cite{cantelar2005tm3+}. 
For example, measurements using a 2\% Tm$^{3+}$-doped LiNbO$_{3}$ crystal revealed $T_1\approx80$ $\mu$s and $T_B\approx 2.4$ ms \cite{cantelar2005tm3+}.

\subsubsection{Structure of Tm$^{3+}$ nuclear-hyperfine levels}
\label{sec:nuclearzeeman}

We apply a magnetic field and perform spectral hole burning to investigate field-activated atomic-level structure and dynamics \cite{moerner1988persistent,liu2006spectroscopic}. 
We expect to observe an atomic-level structure that arises from $^{169}$Tm hyperfine splitting of the $^3$H$_6$ and $^3$H$_4$ levels.
This is due to coupling between the nuclear spin and the enhanced electronic magnetic moment, combined with a weaker nuclear Zeeman effect contribution \cite{sun2012TmLN,thiel2010TmLN}.
Previous measurements of Tm$^{3+}$:Ti$^{4+}$:LiNbO$_{3}$ have shown these levels to have lifetimes of up to several hours at 795.5 nm and 0.85 K with 600 G field \cite{sinclair2017properties}, matching that of bulk Tm$^{3+}$:LiNbO$_{3}$ measured under similar conditions \cite{sinclair2013bulklowT}.

Exposing the Tm-doped crystal at non-zero magnetic field to narrow-band laser light will result in one additional pair of side-holes and up to three pairs of anti-holes in the observed transmission spectrum due to population redistribution among the Tm$^{3+}$ nuclear-hyperfine ground levels \cite{macfarlane1987coherent}.
The depth of the side-holes and anti-holes depends on the measurement timescale and the relative transition rates between hyperfine levels within the excited and ground manifolds. 
However, if strong selection rules prevent optical transitions involving a change in the Tm$^{3+}$ nuclear spin, the spectral hole structure is simplified so that only a single pair of anti-holes appears. 
Furthermore, if magnetic anisotropy or contributions from magneticially-inequivalent sites exist in the crystal, then holes and anti-holes may be broadened or non-resolvable. 
Previous work using bulk Tm$^{3+}$:LiNbO$_{3}$ indicates that several different Tm$^{3+}$ sites may be probed at any given excitation wavelength \cite{thiel2012waveguides,sun2012TmLN}.

We do not immediately resolve anti-holes or side-holes arising from Tm$^{3+}$ hyperfine splitting at a temperature of 0.85 K and field of 600 G for excitation between 794.26 and 795.50 nm wavelengths. 
This is likely due to a strong broadening of the (anti-)hole structure or the high optical absorbance at wavelengths close to 794 nm (see Fig. \ref{fig:inh_echo}). 
Anti-hole broadening has been observed using bulk Tm$^{3+}$:LiNbO$_{3}$ \cite{sun2012TmLN,thiel2012waveguides} as well as bulk and waveguiding Er$^{3+}$:LiNbO$_{3}$ \cite{askarani2019storage, askarani2020persistent}.

\begin{figure}[ht!]
\begin{center}
\includegraphics[width=\columnwidth]{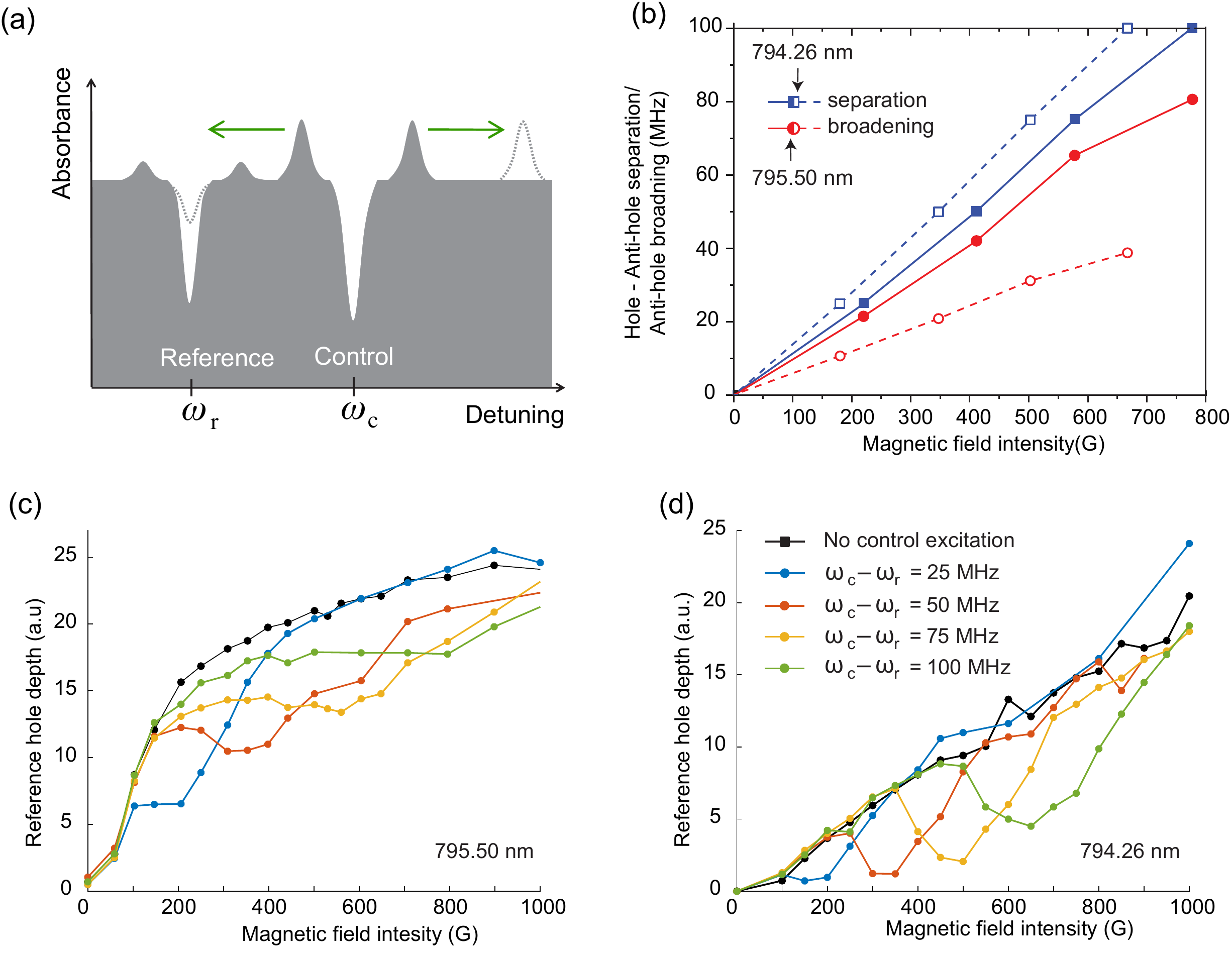}
\caption{Determination of the anti-hole structure due to the Tm$^{3+}$ nuclear-hyperfine interaction. 
(a) Illustration of the concept (see main text for a description).
(b) Anti-hole separations and widths with varied magnetic fields at 795.50 nm and 794.26 nm extracted from data shown in (c) and (d). 
Solid and dashed lines are used to guide the eye.
(c) and (d), Reference hole depths as a function of magnetic field for varying frequency differences ($\omega_c-\omega_r$), at wavelengths of 795.50 nm and 794.26 nm, respectively. 
The legend for (c) is identical to that shown in (d) and solid lines are used to guide the eye. 
} 
 \label{fig:zeeman}
\end{center}
\end{figure}

To determine the anti-hole structure, we perform a modified hole-burning method. 
We simultaneously excite narrow-band subsets of ions at laser frequencies referred to as reference ($\omega_r$) and control ($\omega_c$) and measure the resulting absorbance spectrum at zero magnetic field (i.e. we burn and measure spectral holes at frequencies of $\omega_r$ and $\omega_c$). 
A conceptual example of the absorbance spectrum with non-zero magnetic field depicts exaggerated holes and anti-holes for clarity, where a single pair of anti-holes is ascribed to each hole (Fig. \ref{fig:zeeman}a, grey).

Next, the magnetic field strength is increased to 1 kG while monitoring the shape of the hole at $\omega_r$. 
When the anti-hole generated by the light at $\omega_c$ begins to spectrally overlap with the hole at $\omega_r$, the reference hole depth decreases (Fig. \ref{fig:zeeman}a, dotted line).
Subsequently, the frequency difference between the two holes, $\omega_c-\omega_r$ at minimum hole depth, gives the value of the median separation $\Delta_{sep}$ between the hole and anti-hole.
Repeating this process for varying frequency differences of $\omega_c-\omega_r$ gives rise to anti-hole profiles that broaden with increasing field due to spin inhomogeneous broadening (Fig. \ref{fig:zeeman}c and d for wavelengths of 795.50 nm and 794.26 nm, at 3.5 K and 0.85 K, respectively). 

We determine $\Delta_{sep}$ to vary linearly with field (Fig. \ref{fig:zeeman}b), finding $\Delta_{sep}/2\pi=136 \pm 14$~kHz/G and $155 \pm 6$~kHz/G, for excitation wavelengths of 795.5 and 794.3~nm, respectively.
We attribute $\Delta_{sep}$ to the difference between the hyperfine energy splitting in the $^3$H$_6$ excited level and $^3$H$_4$ ground level.
Our result is consistent with the 140~kHz/G measured directly from field-dependent anti-hole shifts in bulk Tm$^{3+}$:LiNbO$_{3}$ at 794.28~nm and 1.8~K, indicating a similar electronically-enhanced effective nuclear moment in the Tm$^{3+}$:Ti$^{4+}$:LiNbO$_{3}$ waveguide \cite{thiel2010TmLN, sun2012TmLN}.
The differences in $\Delta_{sep}$ with wavelength is attributed to inequivalent Tm$^{3+}$ sites \cite{thiel2012waveguides}.
In addition, our analysis reveals the anti-holes to be Gaussian-shaped, with an inhomogeneous broadening of $\delta\Delta_{sep}/2\pi=109\pm 36~\mathrm{kHz/G}$ and $59\pm 13~\mathrm{kHz/G}$ for measurements at wavelengths of 795.50 nm and 794.26 nm, respectively (Fig. \ref{fig:zeeman}b). 
This indicates a variation of splittings in either the ground or excited state, and similar to what was observed using bulk Tm$^{3+}$:LiNbO$_{3}$, also likely due to inequivalent Tm$^{3+}$ sites \cite{thiel2012waveguides}.
More studies are needed to determine if this broadening can be reduced or how this limits the applicability of Tm$^{3+}$:Ti$^{4+}$:LiNbO$_{3}$ for broadband signal processing.

\subsubsection{Structure of superhyperfine sub-levels}
\label{sec:superhyperfine}

\underline{Spectral hole burning measurements.} 
With increased magnetic fields beyond 1 kG, we expect to resolve more holes and anti-holes due to superhyperfine splitting caused by the weak coupling of the ions of the host crystal to the electronic levels of the REIs \cite{macfarlane1987coherent}.
Superhyperfine ground levels in Tm$^{3+}$:Ti$^{4+}$:LiNbO$_{3}$ can have lifetimes of at least several minutes at 0.85 K for these fields \cite{sinclair2017properties}, similar to those of bulk Tm$^{3+}$:LiNbO$_{3}$ under similar conditions \cite{sinclair2013bulklowT}. 
We probe the superhyperfine structure of the $^3$H$_4$ to $^3$H$_6$ transition using spectral hole burning at a temperature of 0.90 K, with magnetic fields of up to 19 kG, and at wavelengths of 794.26 nm and 795.46 nm.
\begin{figure}[ht!]
\begin{center}
\includegraphics[width=\columnwidth]{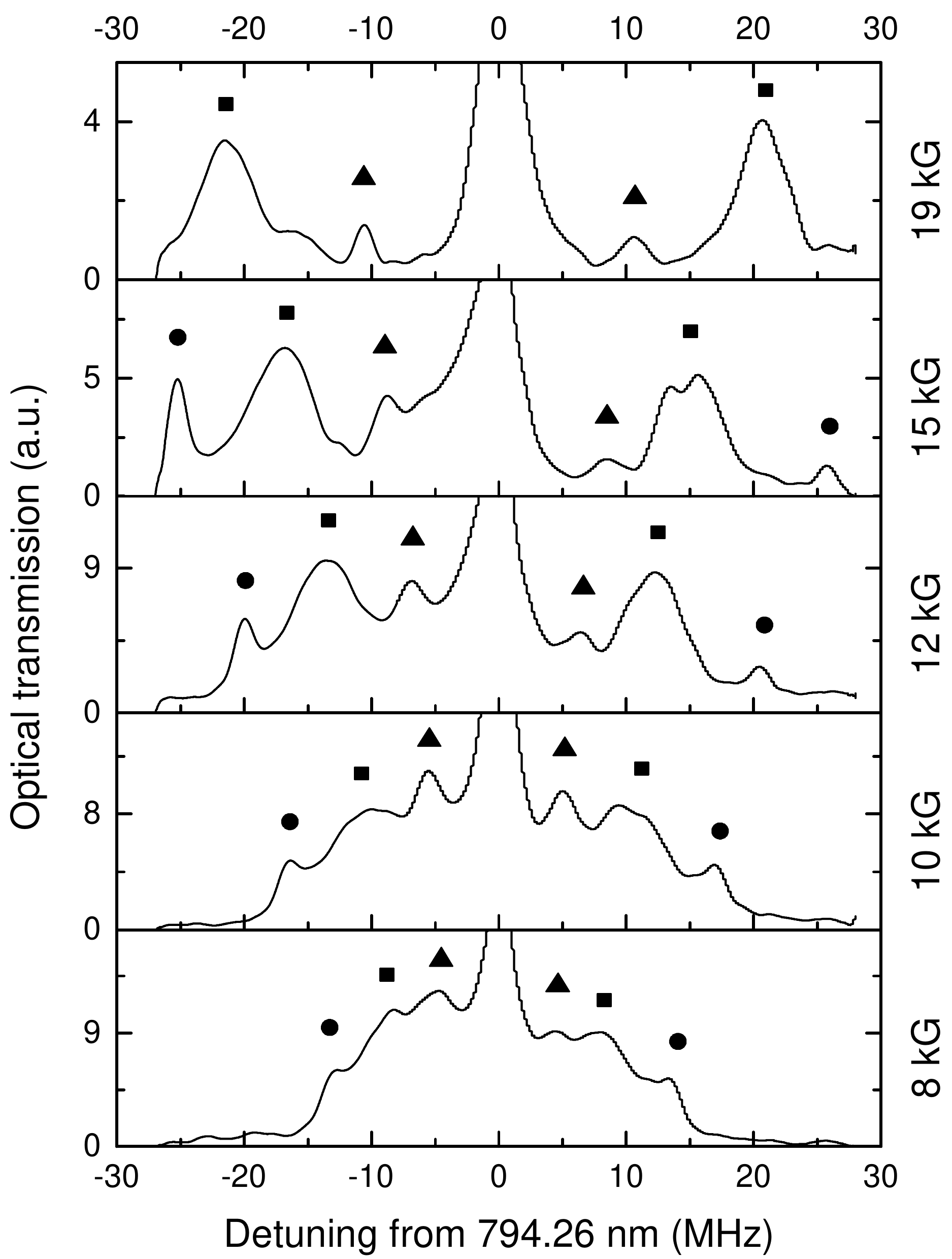}
\caption{Hole burning transmission spectra at a wavelength of 794.26 nm for varying magnetic fields (as indicated).
Triangles, squares, and circles indicate the side holes associated with the $|\Delta m_I|=1$ transitions of $^{6}$Li, $^{93}$Nb, and $^{7}$Li nuclear spins, respectively. 
Distortions are due to the large absorbance.} 
 \label{fig:shf79426}
\end{center}
\end{figure}
\begin{figure}[ht!]
\begin{center}
\includegraphics[width=\columnwidth]{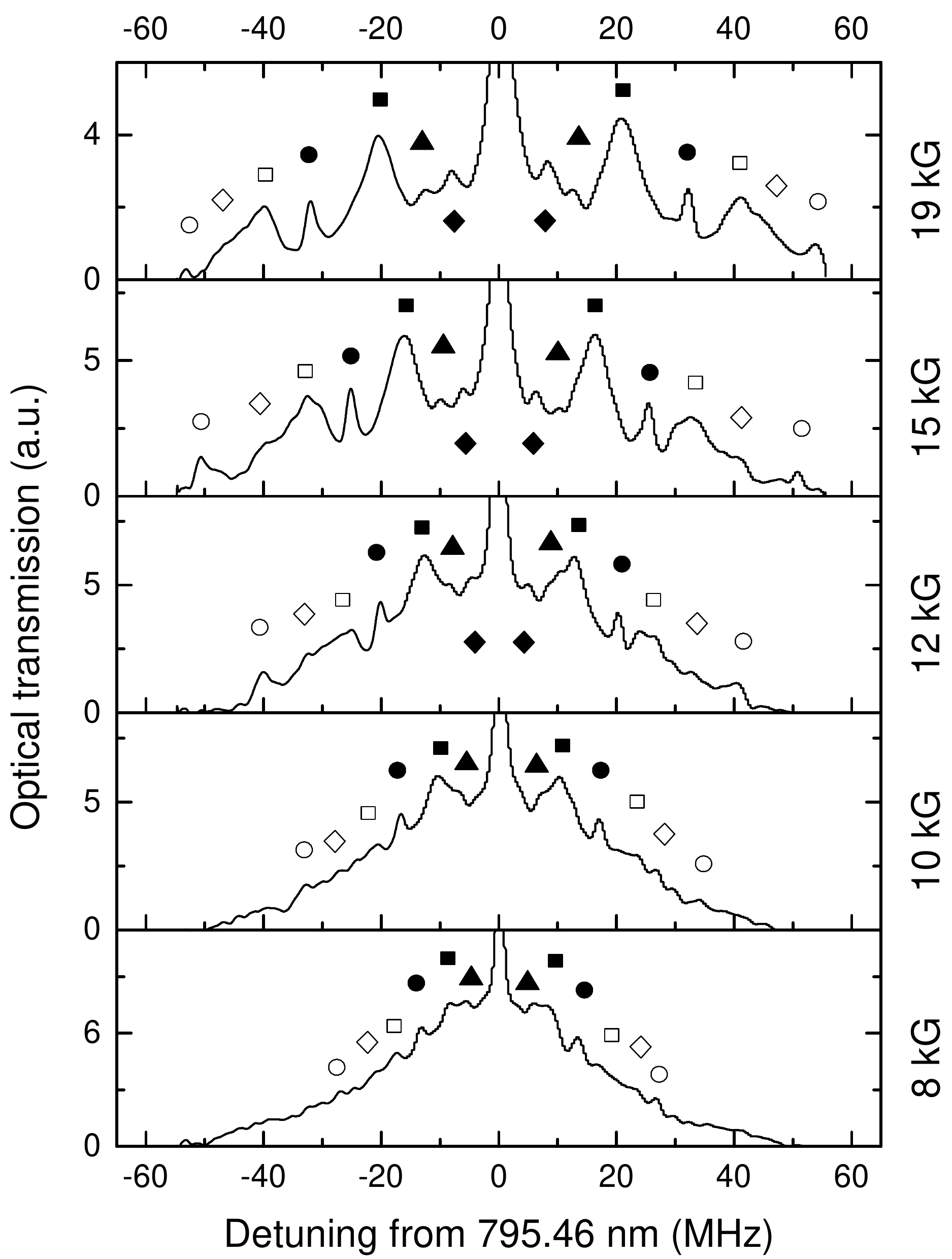}
\caption{Hole burning transmission spectra at a wavelength of 795.46 nm for varying magnetic fields (as indicated).
Triangles, squares, circles, diamonds, open squares, open diamonds, and open circles, indicate the side holes associated with the $|\Delta m_I|=1$ transitions of $^{6}$Li, $^{93}$Nb, and $^{7}$Li nuclear spins, as well as the $|\Delta m_I|=2$ transitions of $^{47,49}$Ti, $^{93}$Nb$_{1}$ (site 1), $^{93}$Nb$_{2}$ (site 2), and $^7$Li nuclear spins, respectively.
Note the increase in the range of detuning compared to Fig. \ref{fig:shf79426}.} 
 \label{fig:shf79546}
\end{center}
\end{figure}

The resultant spectra (Figs. \ref{fig:shf79426} and \ref{fig:shf79546}) reveal no distinct anti-hole structure but rather side-holes due to transitions to different superhyperfine levels in the $^3$H$_4$ excited state \cite{macfarlane1987coherent}.
To identify the spin transitions that are associated with the side-holes, a linear fit of each set of field-dependent side-hole detunings (relative to the main hole) is performed to determine all effective spin gyromagnetic ratios.
We find that the hole structure originates from $^{6,7}$Li, $^{93}$Nb, and $^{47,49}$Ti nuclei in the Ti$^{4+}$:LiNbO$_3$ crystal. 
The ascribed nuclei and transitions, indicated by the difference in magnetic quantum number $|\Delta m_I|$, average gyromagnetic ratios $\gamma$, associated uncertainties from our fits $\delta\gamma$ are indicated in Table I (II) for measurements at a wavelength of 794.26 (795.46) nm.

At 794.26 nm, the splittings are distinctive and match well with those observed using bulk Tm$^{3+}$:LiNbO$_{3}$ at a similar wavelength \cite{thielhbsm2009} despite the presence of more spin inhomogeneous broadening or laser-power broadening (see Sec. \ref{sec:powerdep}). 

At 795.46 nm, the structure is different, likely due to ions experiencing enhanced local strain compared to those probed at 794.26 nm \cite{sun2012TmLN, thiel2012waveguides}. 
Side-holes corresponding to $|\Delta m_I|=1$ transitions of $^{93}$Nb and $^{6,7}$Li have profiles and detunings that are easily identifiable since they compose the two main lattice constituents, and are similar to those observed from measurements using bulk Tm$^{3+}$:LiNbO$_{3}$ \cite{thielhbsm2009} and the results at 794.26 nm. 
The remaining side-holes are attributed to the normally forbidden $|\Delta m_I|=2$ transitions for $^{93}$Nb, $^7$Li, and $^{47,49}$Ti spins. 
Moreover, the $^{93}$Nb sideband is split into two -- likely corresponding to different relative positions to Tm$^{3+}$ ions -- with relative detunings and areas of the two holes giving a center-of-gravity that is similar to the free-ion gyromagnetic ratio. 
Further studies, such as spin double-resonance measurements \cite{kispert1976electron}, are required to fully characterize the nature of the observed superhyperfine splitting and confirm our assignments of spins. 
\begin{table}[ht]
\centering
\begin{tabular}[c]{|l|l|l|l|}
\hline
    element & $\Delta m_I$ & $\gamma$ & $\delta\gamma$ \\
\hline
  $^6$Li & 1 & 0.57 & 0.03 \\
  $^{93}$Nb & 1 & 1.11 & 0.05 \\
  $^7$Li & 1 & 1.71 & 0.01 \\
  \hline
\end{tabular}
\caption{Results of analysis of side-hole structure at varying magnetic fields at a wavelength of 794.26 nm, see main text for details. Units of effective gyromagnetic ratios and associated uncertainties is kHz/G.}
\label{table:794}
\end{table}
\begin{table}[ht]
\centering
\begin{tabular}{|l|l|l|l|}
\hline
  element & $\Delta m_I$ & $\gamma$ & $\delta\gamma$ \\
\hline
  $^{47,49}$Ti & 2 & 0.50 & 0.01 \\
  $^6$Li & 1 & 0.60 & 0.03 \\
  $^{93}$Nb & 1 & 1.04 & 0.01 \\
  $^7$Li & 1 & 1.65 & 0.01 \\
  $^{93}$Nb$_{1}$ & 2 & 1.93 & 0.03 \\
  $^{93}$Nb$_{2}$ & 2 & 2.42 & 0.03  \\
  $^7$Li & 2 & 3.31 & 0.01 \\
\hline
\end{tabular}
\caption{Results of analysis of side-hole structure at varying magnetic fields at a wavelength of 795.46 nm, see main text for details. 
Units of effective gyromagnetic ratios and associated uncertainties is kHz/G.
The subscript on the Nb spin denotes the site associated with its sideband splitting, see main text.}
\label{table:795}
\end{table}

\begin{figure}[ht!]
\begin{center}
\includegraphics[width=\columnwidth]{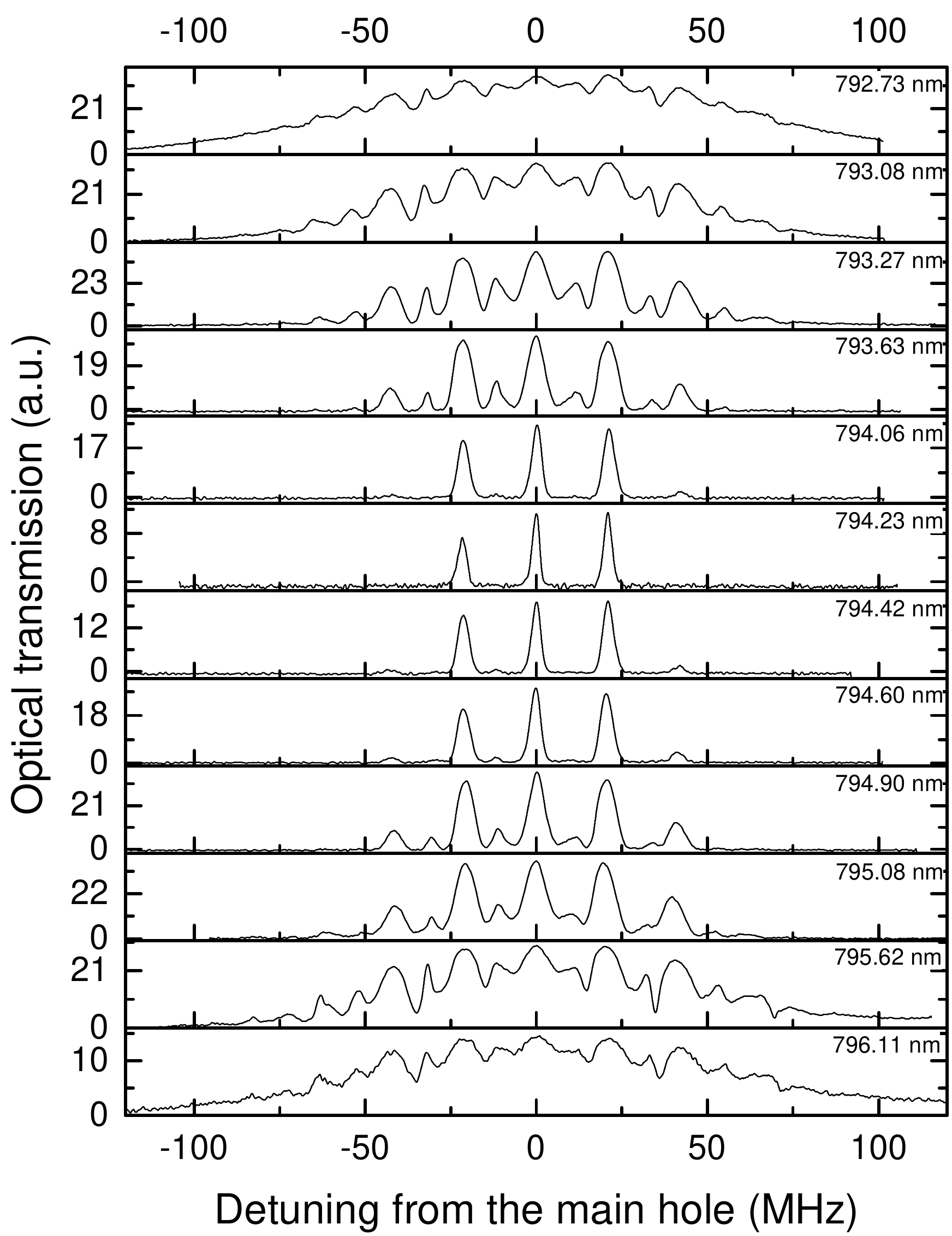}
\caption{Hole burning transmission spectra with 19 kG field for varying laser wavelength (as indicated).
Each measurement is optimized to reduce distortions compared to measurements of Figs. \ref{fig:shf79426} and \ref{fig:shf79546}, yet holes experience greater power broadening at wavelengths where the absorbance is lower.
Vertical scale is optimized for clarity.
}
 \label{fig:shf_multi}
\end{center}
\end{figure}

Next, we set the magnetic field to 19 kG and perform spectral hole burning with varying laser wavelength across the inhomogeneous line to further probe variations of the superhyperfine structure (Fig. \ref{fig:shf_multi}).
We find that the hole structure varies across the inhomogeneous line, with more complex structure at detunings farthest from 794.2 nm, corresponding to Tm$^{3+}$ ions that likely experience higher local strain or are positioned in alternative sites \cite{sun2012TmLN,thiel2012waveguides}. 
At 794.23 nm, the hole spectrum is dominated by the $^{93}$Nb splitting,  similar to that in Fig. \ref{fig:shf79426}.
This is likely due to the strong absorbance that limits observation of the other weaker transitions.
Varying the excitation wavelength towards either 794.9 or 793.6 nm increasingly reveals the splittings from $^6$Li and $^7$Li as well as the $\Delta m_I=2$ transition of $^{93}$Nb.
The side-hole corresponding to $^{47,49}$Ti is hidden, owing to power broadening of the holes.
As the detuning is varied more, either to 796.1 or 792.7 nm (addressing ions experiencing higher strain) more side holes than those in Fig. \ref{fig:shf79546} appear.
Although ascribing transitions to these additional holes is challenging without a measurement with varied magnetic field, we tentatively attribute the split side-hole, corresponding to the $|\Delta m_I|=2$ transition of the $^{93}$Nb ion, to different transitions.
Specifically, we ascribe these two holes to the $|\Delta m_I|=2$ and $|\Delta m_I|=3$ transitions of $^{93}$Nb, acknowledging the limited resolution of Fig. \ref{fig:shf79546}.
This new assignment is consistent with the additional side-holes being from the $|\Delta m_I|=4$ and $|\Delta m_I|=5$ transitions of $^{93}$Nb.
Nevertheless, further studies are required for a complete interpretation of the hole structure.\\

\underline{Photon echo measurements.} If a broad optical inhomogeneous distribution obscures transitions that have small differences in energy splitting, such as the superhyperfine levels in our case, then a two-pulse photon echo decay may feature a modulation due to
the interference between any of the sublevels \cite{macfarlane1987coherent,mims1972envelope,mims1972amplitudes}. 
Similarly, echo modulations may be observed in a three-pulse photon echo decay. 
In the simple case of two doublets with a splitting of $\omega_{g(e)}$ in the ground (excited) state, the system is composed of four optical transitions, leading to a two- or three-pulse photon echo decay described by
\begin{equation}
I(t_{12}) = I_0 e^{-2(\tfrac{2 t_{12}}{T_2} )^x} \big[1 +\frac{\sigma}{1+\sigma^2} F(t_{12},t_{23})\big]^2.
\label{eq:echomod}
\end{equation}
Here the first part is the Mims decay function of Eq. \ref{eq:mims}, $\sigma \equiv \sigma_{diff}/\sigma_{same}$ is the ratio of absorption cross-sections between spins of same or different $m_I$, and $F(t_{12},t_{23})$ is a modulation function that describes the quantum interference between the four possible optical transitions:
\begin{align}
F(t_{12},t_{23}) = e^{-\pi\Gamma_g t_{12}} \{\textrm{cos}[\omega_g t_{12}]+\textrm{cos}[\omega_g (t_{12}+t_{23})]\}\nonumber \\
+e^{-\pi\Gamma_e t_{12}} \{\textrm{cos}[\omega_e t_{12}]+\textrm{cos}[\omega_e (t_{12}+t_{23})]\}\nonumber \\
- e^{-\pi(\Gamma_e + \Gamma_g)t_{12}} \textrm{cos}[\omega_e t_{12}]\textrm{cos}[\omega_g (t_{12}+t_{23})] \nonumber \\
- e^{-\pi(\Gamma_e + \Gamma_g)t_{12}} \textrm{cos}[\omega_g t_{12}]\textrm{cos}[\omega_e (t_{12}+t_{23})].
\label{eq:modulationonly}
\end{align}
Finally, $\Gamma_{g(e)}$ is the nuclear transition linewidths for the ground (excited) states \cite{mims1972envelope,mims1972amplitudes,thiel2010TmLN}. 

\begin{figure}[ht!]
\begin{center}
\includegraphics[width=\columnwidth]{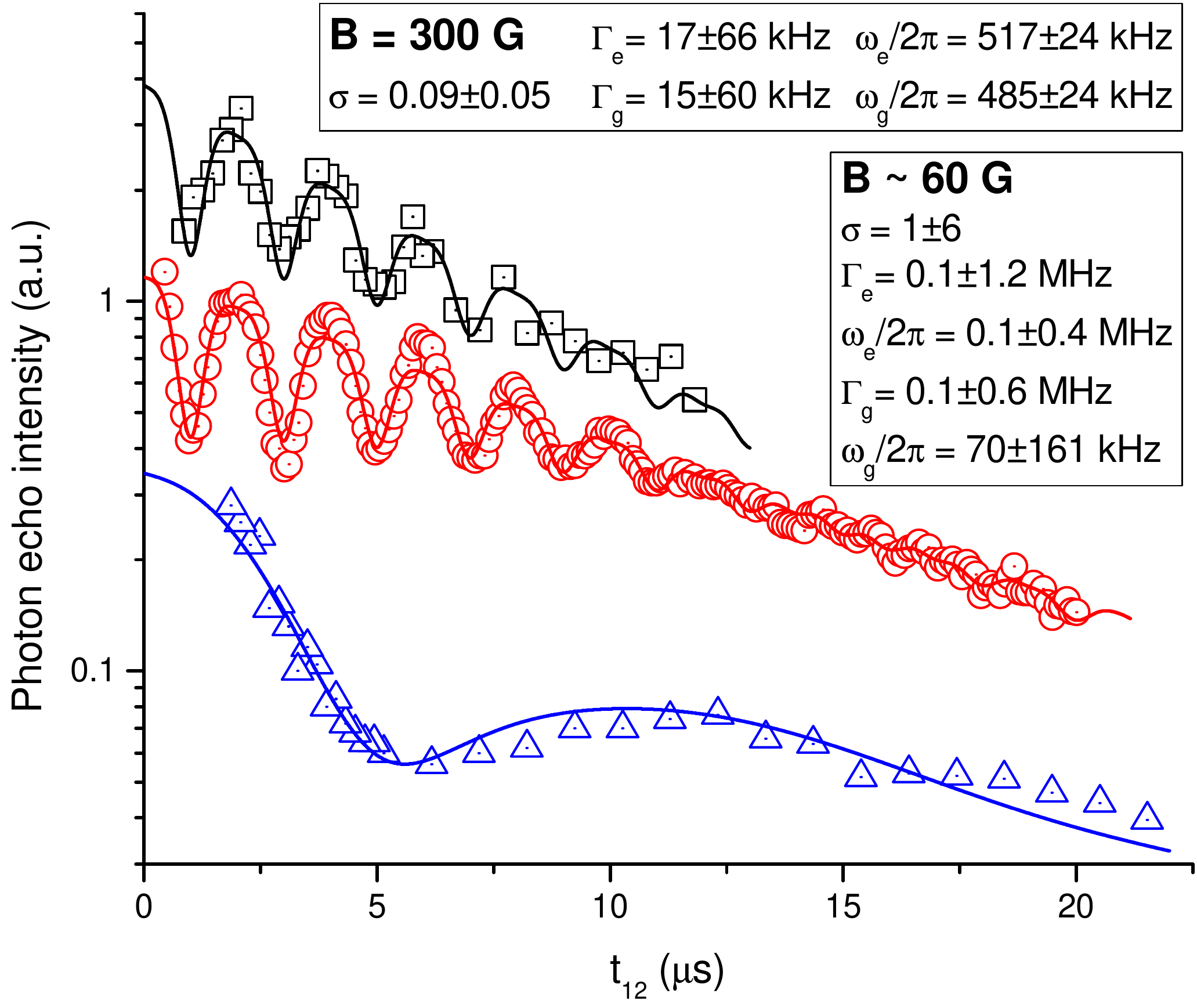}
\caption{Echo decay modulation (log scale) observed using two- (circles, triangles) and three-pulse (squares) photon echo excitation (log scale).
A field of 300 G results in a higher modulation frequency (circles, squares) compared to that obtained using 60 G (triangles).
Parameters extracted from fits using Eq. \ref{eq:echomod} are shown, with large uncertainties due to the large number of fitting parameters or, for the case of the decay with 60 G field, the observation of only one period of modulation.} 
 \label{fig:echomod}
\end{center}
\end{figure}

We perform two- and three-pulse photon echo decay measurements using 300 G field at a wavelength of 795.5 nm and temperature of 0.65 K.
For the latter, $t_{23}$ was fixed at 300 $\mu$s. 
The results are depicted in Fig. \ref{fig:echomod} (circles and squares, respectively). 

We simultaneously fit both decays using Eq. \ref{eq:echomod}.
Our four-transition model describes the observed behavior likely because the $\Delta m_I>1$ transitions are weak and any nuclear quadrupole splitting is very small \cite{thiel2010TmLN}.
The fit reveals modulation frequencies of $\omega_{g(e)}/2\pi=$ 485$\pm$ 24 (517$\pm$24) kHz that, considering the results presented in Table II, correspond to the $|\Delta m_I|=1$ splitting of the $^7$Li nucleus (495 kHz) with a nuclear quadrupole shift of 22 kHz \cite{thiel2010TmLN}. 
Linewidths $\Gamma_{g(e)}=$ 15$\pm$60 (17$\pm$66) kHz and a cross-section ratio of 0.09$\pm$0.05 are extracted from the fits. 
All parameters extracted from the fit are consistent with $\omega_{g(e)}/2\pi=$ 498 (520) kHz, $\Gamma_g=\Gamma_e=$ 18 kHz, and $\sigma=0.06$, measured using a Tm$^{3+}$:LiNbO$_{3}$ bulk crystal at 1.8 K, a wavelength of 794.26 nm, and the same field \cite{thiel2010TmLN,thielhbsm2009}, which was also ascribed to the same $^7$Li transition.

Finally, to verify the field-dependence of the modulation, we reduce the magnetic field to $\sim$60 G, and measure a two-pulse photon echo decay shown in Fig. \ref{fig:echomod} using triangles.
For these measurements, the temperature is 0.8 K.
A fit using Eq. \ref{eq:echomod} yields $\omega_{g(e)}/2\pi=$ 70$\pm$161 (100$\pm$400) kHz, which is consistent with the measurements using 300 G field. 
Note that we do not observe a modulation due to any other superhyperfine transition, consistent with the observation using bulk Tm$^{3+}$:LiNbO$_{3}$ at 794.26 nm.
This may be due to similar superhyperfine splittings from different $\Delta m_I$ transitions of $^7$Li in the $^3$H$_6$ and $^3$H$_4$ levels, or weak emission from levels split by other nuclear spins \cite{thiel2010TmLN,thielhbsm2009}.
\\ \\
\underline{Measurements using continuous coherent excitation.} Superhyperfine structure may also be revealed by continuous coherent narrow-band excitation of the optical transition.
Due to inhomogeneous broadening, this results in coherent emission by optical transitions that differ in energies given by the superhyperfine splittings.
This produces a modulation of the transmitted field, sometimes referred to as a quantum beat \cite{haroche1973quantumbeat}.
\begin{figure}[ht!]
\begin{center}
\includegraphics[width=\columnwidth]{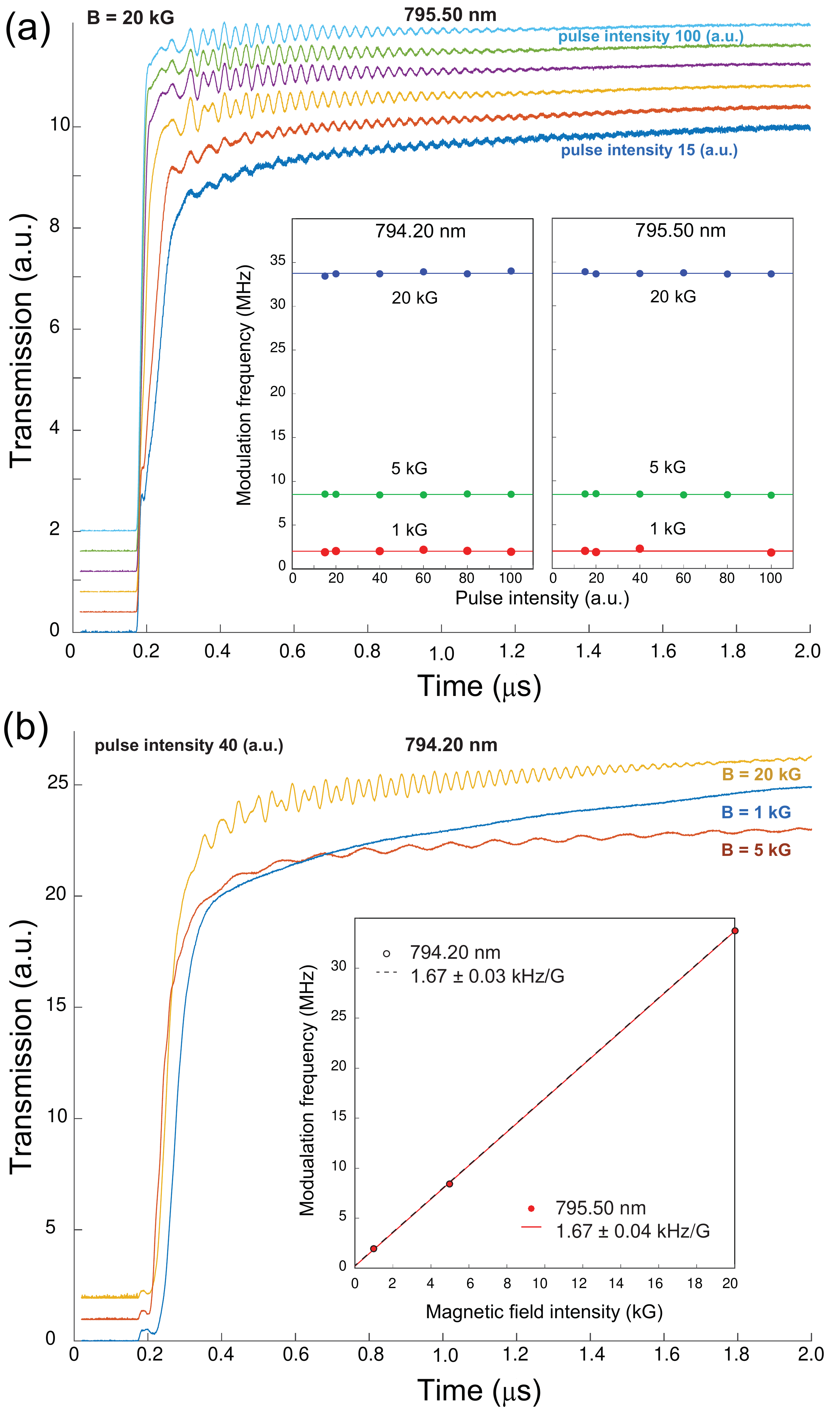}
\caption{Intensity modulation of a narrowband pulse of light caused by superhyperfine levels.
(a) The transmitted intensity of pulses with varying input powers (vertically displaced) at a wavelength of 795.50~nm and a magnetic field of 20~kG. 
The inset depicts the most discernible modulation frequencies and their dependence on magnetic field, wavelength, and the input pulse intensity.
(b) The transmitted intensity for pulses of fixed intensities with varying magnetic field at a wavelength of 794.20~nm. 
The inset shows that the frequencies of the most discernible modulation varies linearly with magnetic field and is independent of wavelength.
Fits, represented by dotted lines, reveal the modulation to be from $^7$Li superhyperfine splitting.
} 
 \label{fig:pulsemod}
\end{center}
\end{figure}

To this end, for two different laser excitation wavelengths, three different magnetic fields, and using varying optical excitation power, we observe the modulation of a long optical pulse after being transmitted through the waveguide.
These measurements are performed at 1 K.
Figs.~\ref{fig:pulsemod}a and insets show that the modulation is similar for all optical powers.
This modulation cannot be related to optical nutation, which is the coherent driving, and hence modulation, of atomic population between optical transitions, which has a rate that is proportional to the square root of the excitation power.
Further, we observe the amplitude and periodicity of the modulation at the same magnetic field to be similar for both wavelengths (Figs.~\ref{fig:pulsemod}a and b).
As the field is increased, the modulation becomes damped, while the modulation frequency increases and becomes more complex.
As shown in the inset of Fig.~\ref{fig:pulsemod}b, fits of the most discernible modulation frequency for both excitation wavelengths reveal a linear and identical dependence with respect to the magnetic field, with a slope of $1.67\pm 0.04$~kHz/G and zero offset (within error).
The modulation frequency is similar to that observed using photon echoes, and, according to Tables I and II, consistent with the splitting caused by coupling to $^7$Li.
Moreover, the weak intensity of the modulation is consistent with the small absorption cross-sections predicted by the echo data.
The damping of the modulation is due to spin inhomogeneous broadening, occurs faster for higher magnetic fields, and is more pronounced at a wavelength of 795.5 nm, consistent with the broadening of the nuclear-hyperfine levels.
The additional frequency components of the modulation at high fields could be due to the similar $\Delta m_I$ transitions caused by $^7$Li in the $^3$H$_6$ and $^3$H$_4$ levels.
However, more measurements are required to evaluate this interpretation.

\subsubsection{Excitation-power dependence of spectral hole widths and depths}
\label{sec:powerdep}

Spectral tailoring of an optical inhomogeneous line is key to realizing optical applications with REICs such as optical filters or dispersion elements \cite{moerner1988persistent,thiel2011applications}. 
Narrow-band excitation will produce a spectral hole whose depth and width depend not only on the level structure and dynamics, but also on the intensity and the duration of the excitation pulse \cite{moerner1988persistent}. 
Here we investigate the effects of laser power on the widths and depths of spectral holes using the Tm$^{3+}$:Ti$^{4+}$:LiNbO$_{3}$ waveguide.
\\ \\
\underline{Power dependence of hole width.} Strong laser powers can rapidly drive transitions. 
This leads to laser-induced broadening of a spectral hole, an effect referred to as power broadening \cite{maniloff1995power}. 
To measure power broadening, we perform spectral hole burning at 0.85 K, zero magnetic field, and 795.5 nm wavelength with varying laser power.

\begin{figure}[ht!]
\begin{center}
\includegraphics[width=\columnwidth]{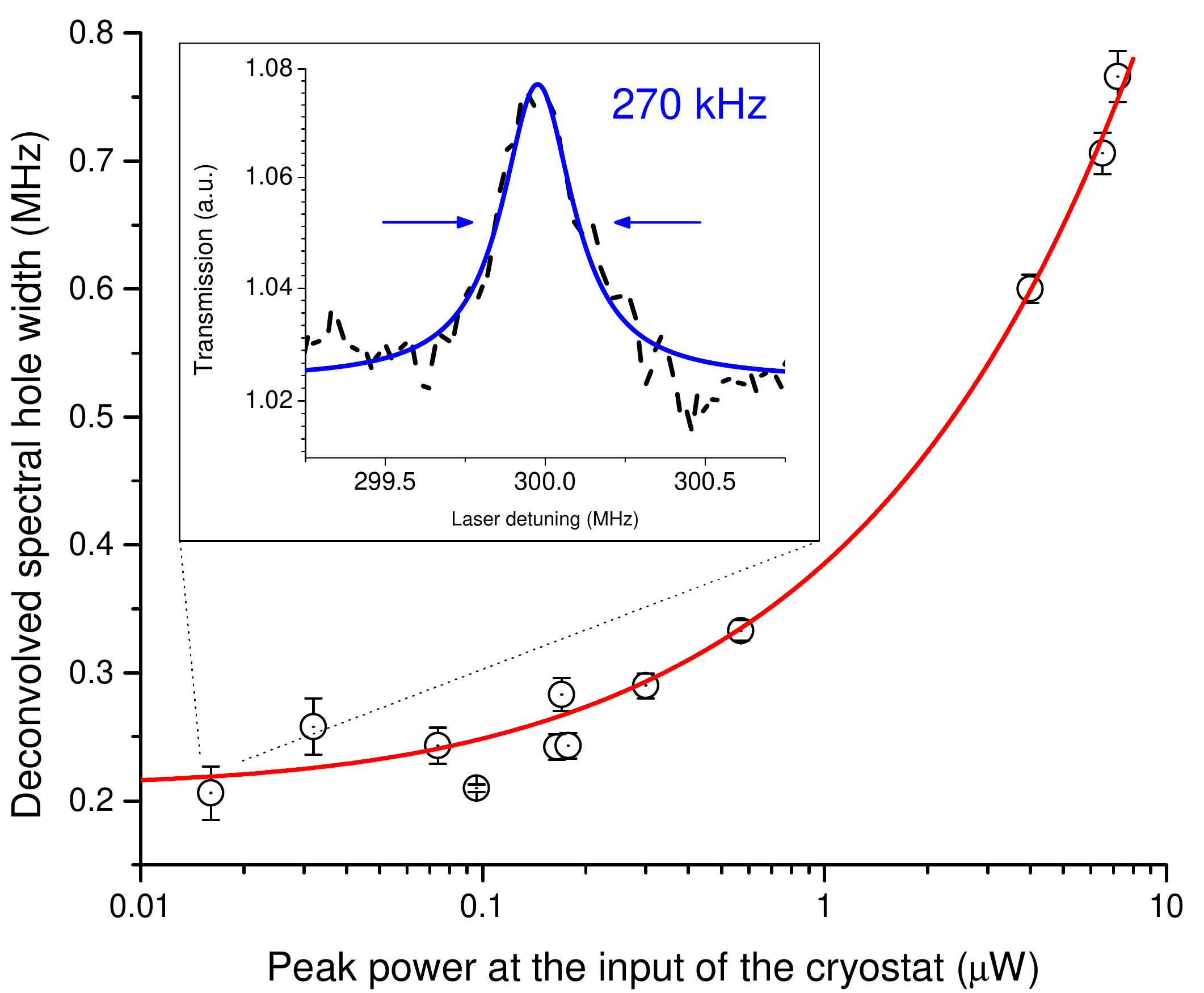}
\caption{Power broadening of a spectral hole. 
Linear-log scale is used to clearly reveal the zero-excitation intercept. 
The hole width is deconvolved from 70 kHz of broadening due to the laser frequency variation. 
Inset: Spectral hole and fit of the data taken using 16 nW of excitation power.} 
 \label{fig:pb}
\end{center}
\end{figure}

Ions are excited using a pulse of 1 ms duration and, by varying the laser detuning after a delay of 400~$\mu$s, the shape of the spectral hole is measured. 
The full-width-at-half-maximum width $\Gamma_{hole}$ of the hole is fit using a Lorentzian (Fig. \ref{fig:pb}).
For excitation powers of $< 100$ nW, we observe no power dependence of the hole width, indicating that our narrowest spectral hole width is limited by laser frequency jitter and spectral diffusion.
Power broadening increases with excitation power according to
\begin{align}
\Gamma_{hole} = \Gamma_{L}\big[(1+\sqrt{1+(\Gamma_{L} K)^2})\nonumber \\
\times (1+\sqrt{1+(\Gamma_{L}K)^2 e^{-d{_i}}})\big]^2,
\label{eq:pb}
\end{align}
where the initial absorbance of the transition is $d_i$, $\Gamma_{L}$ is the fitted homogeneous linewidth that includes linewidth broadening due to laser frequency instability, and $K^2$ is proportional to the excitation power \cite{maniloff1995power}. 
We fit our data using Eq. \ref{eq:pb}, yielding $d_i=2.9\pm 1.3$ ( which is comparable to the $d=2.4$ from Fig. \ref{fig:inh_echo}) and $\Gamma_{L}=106 \pm 4$ kHz, which, given the $<$100 kHz laser linewidth, is compatible with the $\sim50$ kHz measured using bulk Tm$^{3+}$:LiNbO$_{3}$ at 1.7 K and 794.27 nm wavelength \cite{thiel2010TmLN,sun2012TmLN,thiel2014ISD}.

\underline{Power dependence of hole depth.} It is a recurrent observation when spectrally tailoring Tm$^{3+}$:Ti$^{4+}$:LiNbO$_{3}$ that wide spectral features may not be burned to full transparency.
This may be due to accumulation of population in near-detuned inhomogeneously broadened hyperfine levels, which results in a reduction of hole depth, or ISD that also results in hole broadening and depth reduction \cite{thiel2014ISD}.
However, as described below, we rule out all of these causes and attribute the limitation to spin-lattice relaxation that is induced by laser excitation.

\begin{figure}[ht!]
  \centering
  \includegraphics[width=\columnwidth]{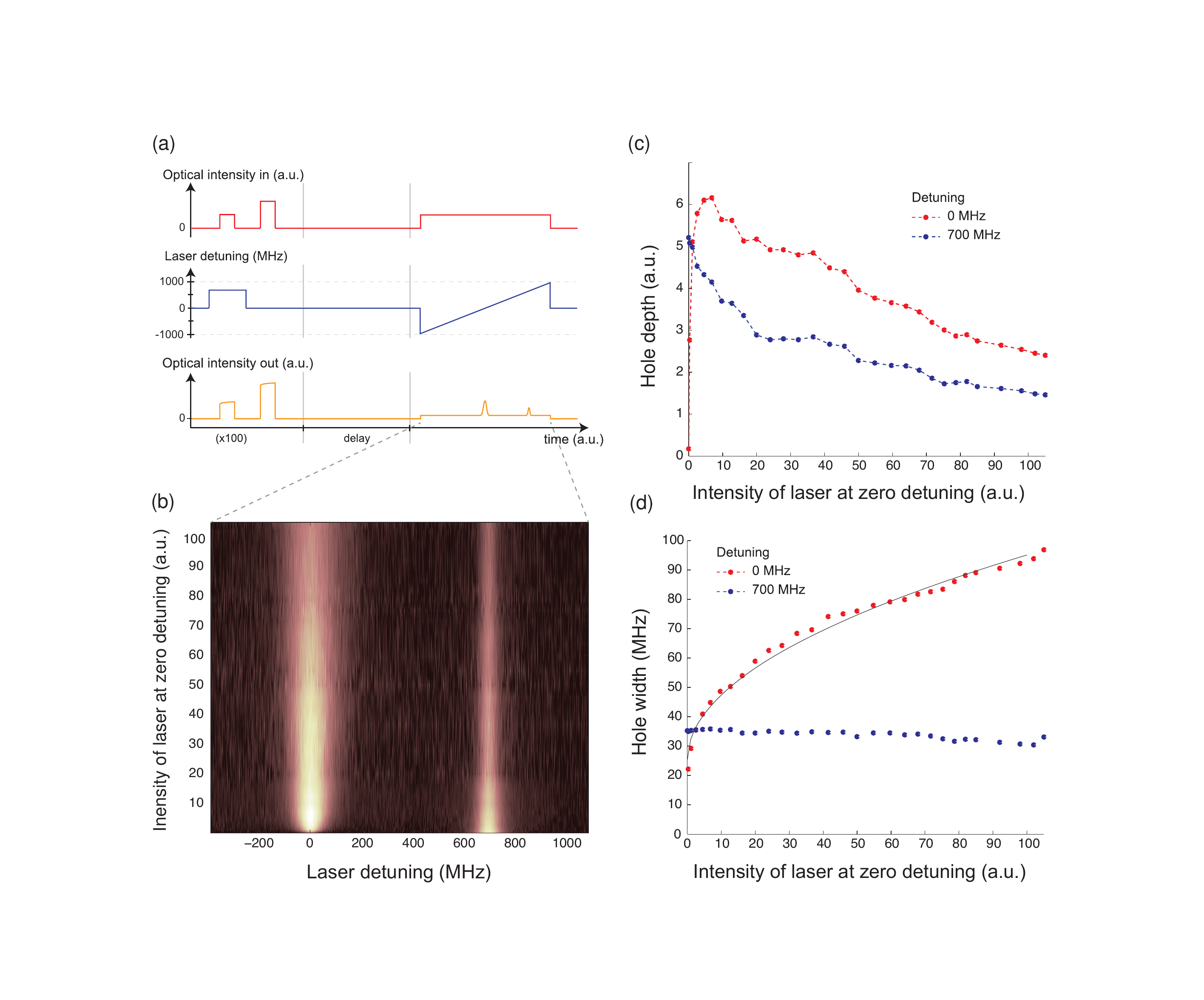}
\caption{Excitation power-dependence of the depth of a spectral hole. 
(a) Timing sequence of the experiment. Optical intensity in (out) indicates the light that is generated (detected) before (after) the waveguide.
(b) Hole burning absorbance spectrum for varying excitation powers at zero detuning. 
Dark (light) color corresponds to high (low) absorbance. 
(c) Depths of both holes are reduced with increased burning power.
(d) Width of hole at 700 MHz detuning remains nearly constant, while the other is power broadened.
} 
\label{fig:powdep_holedepth}
\end{figure}

To investigate this effect, we perform the following experiment using a 3~kG field at 3~K and a wavelength of 795.5~nm (Fig.~\ref{fig:powdep_holedepth}a).
We consecutively burn two spectral holes, first one at 700~MHz detuning using a fixed excitation power and then one at 0~MHz detuning with varying excitation power, starting with none. 
The detuning avoids the impact of anti-holes from population accumulation in nuclear-hyperfine levels. 
After a 2~ms delay, we record the optical absorbance profile and fit the hole depths and widths at the two detunings with varying excitation power used to burn the 0~MHz detuned hole (Figs.~\ref{fig:powdep_holedepth}b, c, d).

As expected, we observe that the depth of the hole at 0~MHz initially increases as the increased power allows more ions to be optically pumped. 
However, a further increase in optical power results in a decrease in hole depth.
Note that the width of this hole increases with the square-root of the power, as expected from power broadening. 
Meanwhile, the hole at 700~MHz detuning decreases as soon as optical power is applied to burn the 0~MHz hole.
Since the power employed to burn the 700~MHz hole is constant, we do not expect any power broadening of that hole. 
Specifically, we observe that the width of the hole at 700~MHz detuning is constant with excitation power at 0~MHz detuning. 
Since the area of the hole at 700~MHz detuning is not conserved, we cannot ascribe the reduction of this hole to ISD.
Although its width is increased by power broadening, we expect that the reduction of the hole at 0~MHz detuning also cannot be explained by ISD. 

We note that this measurement procedure has also been performed using an Er$^{3+}$:Ti$^{4+}$:LiNbO$_{3}$ waveguide, with similar conclusions (i.e. ruling out population redistribution in hyperfine levels and ISD) \cite{askarani2020persistent}.
However, excitation-power dependence of hole depth has not been observed in bulk REI-doped LiNbO$_{3}$ \cite{thiel2019private}, likely because its effect is significantly enhanced by light confinement in a waveguide.
%Moreover, we do not expect any effects due to magnetic dipole-dipole interactions that are driven by laser excitation, e.g. due to decay of Tm$^{3+}$ from the $^3$H$_4$ excited level (see Sec. \ref{sec:spectraldiffusionwithfield}), given the strong magnetic field used for this measurement \cite{thiel2012waveguides,thiel2014ISD}.
Following the discussion in Ref. \cite{askarani2020persistent}, the reduction in hole depth is furthermore not caused by accelerated spin diffusion, i.e. spin flip-flops due to nuclear spin excitation and decay during the burning procedure, and likely not by coupling of laser-excited two-level tunneling systems (TLSs) to nuclear spins as we have not observed any clear signature of TLS in any of our previous (coherence) measurements.
Alternatively, spin-lattice relaxation from non-equilibrium phonon dynamics during the laser excitation \cite{liu2006spectroscopic, graf1998phonon}, in which phonons are generated due to electron-phonon interactions and the decay of optically-excited levels of REIs, could play a role.
Nonetheless, further measurements are needed to clarify the presence of these dynamics, (e.g. using varying laser detuning and magnetic fields) or using laser excitation that is far detuned from the REI transition to determine if the effects are caused by the Ti$^{4+}$:LiNbO$_{3}$ crystal itself.

\section{Conclusion}
\label{sec:conclusion}

We analyzed the coherence and energy-level properties of the $^3$H$_6$ to $^3$H$_4$ optical transition of a Tm$^{3+}$:Ti$^{4+}$:LiNbO$_{3}$ waveguide at temperatures as low as 0.65 K, with fields up to 20 kG, over varying measurement timescales as well as laser excitation wavelengths and powers.
Complementing our previous work, we characterize properties that limit the performance of this material for optical signal processing applications, shedding light on underlying mechanisms behind important parameters such as optical coherence or hole spectra.

Importantly, we find properties that are consistent with those of a Tm$^{3+}$:LiNbO$_{3}$ bulk crystal at temperatures of less than 1 K except for differences that can be explained by Tm$^{3+}$ or Ti$^{4+}$-doping concentration.
For example, Tm$^{3+}$-doping appears to impact the temperature-dependence of homogeneous linewidth and reduces the excited-level lifetime compared to that measured in the bulk crystal, while mode confinement provided by Ti$^{4+}$-doping yields additional side-holes while facilitating strong laser intensities that enhance the spin-lattice relaxation rate of the hyperfine levels. 
Still, some properties could not be compared because they were not measured in the bulk crystal, e.g. time-dependent spectral diffusion using a 20 kG field or wavelength-dependent superhyperfine structure.

Our study indicates that REIs retain their properties when the crystal is co-doped for integrated applications, thereby establishing new directions for optical signal processing in this widely-used electro-optic crystal.

\section{Acknowledgments}
We thank M. George, R. Ricken and W. Sohler for fabricating the waveguide, and to M. Hedges, H. Mallahzadeh, T. Lutz, L. Veissier, C. Deshmukh, and M. Falamarzi Askarani for discussions. 
We acknowledge funding through the Natural Sciences and Engineering Research Council of Canada (NSERC), Alberta Ministry for Jobs, Economy and Innovation's Major Innovation Fund on Quantum Technology, Alberta Innovates Technology Futures (AITF) research program, Defense Advanced Research Projects Agency (DARPA) Quiness program (Contract No. W31P4Q-13-1-0004), National Science Foundation (NSF) under award nos. PHY-1415628 and CHE-1416454, NSF Science and Technology  ``Center for Integrated Quantum Materials" under Cooperative Agreement No. DMR-1231319, Department of Energy/High Energy Physics QuantISED program grant, QCCFP (Quantum Communication Channels for Fundamental Physics), award number DE-SC0019219, AQT Intelligent Quantum Networks and Technologies (INQNET) research program, and the Dutch Research Council (NWO).

%\bibliography{TmLNbib}
%removed bib file using bbl

\end{document}